\documentclass[a4paper,fleqn,usenatbib]{mnras}
\usepackage{xcolor}

\usepackage{latexsym,amssymb,amsmath,esint,graphics,graphicx,soul} 

\def\l{\bigg{(}} \def\r{\bigg{)}} \def\la{\langle} \def\ra{\rangle}
\def\d{\mathrm{d}} \def\v{\mathbf{v}} 
\def\V{\boldsymbol{\mathcal{V}}}  \def\E{\mathbf{E}}
\def\B{\mathbf{B}}   \def\M{\mathbf{M}}
\def\F{\mathbf{F}} \def\G{\mathbf{G}} \def\H{\mathbf{H}} \def\U{\mathbf{U}}
 \def\Pgas{P_\mathrm{gas}}
\def\mathF{\mathcal{F}}
\def\div{\vec{\nabla} \cdot}
\newcommand{\der}[2]{\frac{\mathrm{d}#1}{\mathrm{d}#2}}
\newcommand{\pder}[2]{\frac{\partial #1}{\partial #2}}
\newcommand{\dpder}[2]{\dfrac{\partial #1}{\partial #2}}
\def\nui{{\nu_i}}
\def\msun{M_\odot}

\title[MHD simulations of NS merger accretion disks]{Long-term 3D-MHD 
       Simulations of Black Hole Accretion Disks formed in Neutron Star Mergers}

\author[Fahlman \& Fern\'andez]{
Steven Fahlman$^{1}$\thanks{E-mail: sfahlman@ualberta.ca}, Rodrigo Fern\'andez$^{1}$\thanks{E-mail: rafernan@ualberta.ca}\\
$^{1}$Department of Physics, University of Alberta, Edmonton, AB T6G 2E1, Canada\\
}

\pubyear{2022}

\begin{document}
\label{firstpage}
\pagerange{\pageref{firstpage}--\pageref{lastpage}}
\maketitle

\begin{abstract} 
We examine the long-term evolution of accretion tori around black hole (BH)
remnants of compact object mergers involving at least one neutron star, to
better understand their contribution to kilonovae  and the synthesis of
r-process elements.  To this end, we modify the unsplit
magnetohydrodynamic (MHD) solver in \texttt{FLASH4.5} to work in
non-uniform three-dimensional spherical coordinates, enabling more
efficient coverage of a large dynamic range in length scales while 
exploiting symmetries in the system.  This 
modified code is used to perform BH accretion disk
simulations that vary the initial magnetic field geometry and disk compactness,
utilizing a physical equation of state, a neutrino leakage scheme for
emission and absorption, and modeling the BH's gravity with a
pseudo-Newtonian potential.  Simulations run for long enough to achieve a
radiatively-inefficient state in the disk.  We find robust mass
ejection with both poloidal and toroidal initial field geometries, and
suppressed outflow at high disk compactness. With the included physics, we
obtain bimodal velocity distributions that trace back to mass ejection
by magnetic stresses at early times, and to thermal processes in the
radiatively-inefficient state at late times. The electron fraction distribution
of the disk outflow is broad in all models, and the
ejecta geometry follows a characteristic hourglass shape.  We test the effect
of removing neutrino absorption or nuclear recombination with axisymmetric
models, finding $\sim 50\%$ less mass ejection and more neutron-rich composition
without neutrino absorption, and a subdominant contribution from nuclear
recombination. Tests of the MHD and neutrino leakage implementations are
included.
\end{abstract}

\begin{keywords}
accretion, accretion disks -- MHD -- neutrinos -- nuclear reactions, nucleosynthesis, abundances 
           --- stars: black holes --- stars:neutron
\end{keywords}

\section{Introduction} \label{sec:intro}

Neutron star (NS) mergers have long been predicted to produce $r$-process
elements  \citep{Lattimer1974} as well as electromagnetic (EM) transients
powered by relativistic jets \citep{Paczynski86, Eichler+89} and/or by the
radioactive decay of newly formed heavy elements \citep{LP1998,Metzger2010}.
The detection of a short gamma-ray burst (SGRB) and a kilonova associated with
the NS merger GW170817 \citep{ligo_gw170817_gw,ligogw170817multi-messenger} has
confirmed these events as progenitors of SGRBs, and placed them as important
production sites for the $r$-process given the broad agreement between kilonova
observations and predictions (e.g., \citealt{Cowperthwaite2017,chornock2017,
drout2017,Tanaka2017,tanvir2017}).

Despite the broad agreement with observations, key properties
of the kilonova from GW170817 have not yet been fully accounted for theoretically.
Observations require a multi-component signal including (at least) a fast blue transient 
and a slower red component containing most of the ejecta mass 
(e.g., \citealt{Kasen2017,Villar2017}). The red component had been predicted based on the 
high opacity of lanthanide-rich material \citep{Kasen2013,Tanaka2013,Barnes2013,Fontes2015}
and is the crucial piece linking NS mergers to $r$-process nucleosynthesis.
A blue component had been anticipated as a possible consequence of a finite-lived
hypermassive NS (HMNS) irradiating the ejecta with neutrinos \citep{MF14}, but 
the low amount of dynamical ejecta expected (e.g., \citealt{Shibata2017,most_2019}) and the low velocities
of HMNS disk outflows in viscous hydrodynamics
\citep{FahlmanFernandez2018,Vsevolod2020} rule out a 
straightforward interpretation. Remaining explanations include magnetically-driven
outflows (e.g., \citealt{Metzger2018}), opacity effects (e.g., \citealt{Waxman2018}), 
or connection to a jet cocoon (e.g., \citealt{gottlieb_2018,piro_2018}). 

More generally, the third observing run of Advanced LIGO \& Virgo
\citep{ligo_gwtc-2,ligo_gwtc-3} found NS-NS and/or NS-black hole (BH) candidates
for which no EM counterpart was detected.
The lack of counterparts can be explained by either observational factors (e.g.,
\citealt{foley_2020}), smaller amounts of mass ejected than in GW170817 given
different possible binary properties, or a combination of the two. Better
understanding the physics of mass ejection is therefore of key importance to
improve predictions and understanding of future detections (e.g.,
\citealt{raaijmakers_2021}).

Mass ejection from the accretion disk formed during a NS merger can be
comparable or even dominate over material ejected dynamically depending on the
binary properties (e.g., \citealt{Radice2018Ensemble,Kruger2020,Vsevolod2020}).
Disk outflows are driven by a combination of neutrino absorption
\citep{ruffert1997}, magnetic stresses (e.g., \citealt{devilliers_2005}), and
thermal processes \citep{metzger2009_freezout}, over timescales ranging from the
orbital time ($\sim $ few ms) to the angular momentum transport timescale ($\sim
0.1-1\,$s).  Several studies in axisymmetric viscous hydrodynamics have
characterized the amount and composition of the outflow over the longest
timescales in the problem, finding that a significant fraction of the initial
disk mass is ejected as a neutron-rich outflow that can generate $r$-process
elements and contribute to the kilonova transient (e.g.,
\citealt{FM13,Just2015,Fujibayashi2020BHNS}). A central HMNS can increase the
fraction of the disk ejected relative to a BH, with a less neutron-rich
composition due to more intense neutrino irradiation
\citep{MF14,Perego2014,Lippuner2017,FahlmanFernandez2018,fujibayashi2018,
Fujibayashi2020,Shibata2021MeanField,curtis_2021}. 

It is generally accepted, however, that angular momentum transport in
astrophysical disks is mediated by the magnetorotational instability
\citep[MRI,][]{Hawley1991}.  Capturing this instability requires
three-dimensional (3D) magnetohydrodynamic (MHD) simulations, which are
computationally expensive given the long evolutionary times in the problem.
Consequently, only a handful of simulations of BH accretion disks in 3D
general-relativistic (GR) MHD with some form of neutrino physics have been
conducted thus far
\citep{nouri_2017,siegel_2017a,Siegel2018,Fernandez2019,Miller2019,Christie2019,just_2021,
murguia-berthier_2021,Hayashi2021} following earlier work in axisymmetry
\citep{Shibata2007,Shibata2012,janiuk2013,janiuk_2017}. 

These works show that the physics of late-time mass ejection in GRMHD BH
accretion disks are broadly consistent with viscous hydrodynamics.  Thermal mass
ejection due to dissipation of MRI turbulence following the freezeout of weak
interactions, along with nuclear recombination, behave in a similar manner as the
analog process in viscous hydrodynamics \citep{Fernandez2019}, but magnetic
stresses can enhance mass ejection significantly, with a much broader
distribution in electron fraction. This results from the difference between the
viscous and magnetic mass ejection timescales, and is dependent on the initial
field strength and configuration (e.g., \citealt{Christie2019}).

Disks around HMNSs in MHD are a more 
challenging problem, and only limited exploration for short times
has thus far been done in 3D 
(e.g., \citealt{Kiuchi2012,Siegel_2014,Ciolfi2020,moesta_2020}, see
also \citealt{shibata_2021,Shibata2021MeanField} for long-term evolution in 2D).

Here we introduce a relatively inexpensive computational approach for exploring
long-term mass ejection from BH accretion disks by solving the MHD equations
with a pseudo-Newtonian potential to model the BH, and a neutrino leakage scheme
that includes both emission and absorption. We use this method to explore the
role of magnetic field geometry, disk compactness, nuclear recombination, and
neutrino absorption on mass ejection in MHD. We perform simulations for initial
conditions relevant to GW170817, as well as to systems that could feasibly
result from a NS-BH merger. The main limitation of our approach is the absence
of relativistic jets, thus our focus is on the sub-relativistic outflows that
contain most of the ejected mass, and which are therefore most relevant to the kilonova
emission and $r$-process nucleosynthesis.

The structure of this paper is the following. Section \S\ref{sec:methods} presents
a description of the numerical methods employed and models evolved. In
\S\ref{sec:results} 
the results of our simulations are presented, analyzed, and compared
to previous work.
We conclude and summarize in
\S\ref{sec:conc}. The appendices
describe the testing and implementation of the MHD
{(\S\ref{AppA}, \S\ref{AppB})} and neutrino leakage
{(\S\ref{AppC})} modules employed.

\section{Methods} \label{sec:methods}

\subsection{Numerical MHD}
Our simulations employ a customized version of \texttt{FLASH} version 4.5
\citep{Fryxell2000,Dubey2009}, in which we have modified the unsplit MHD solver
of \citet{Lee2013} to work in 3D curvilinear coordinates with non-uniform
spacing  (see Appendix~\ref{AppA} for details). We use this code to numerically
solve the Newtonian equations of mass, momentum, energy, and lepton number
conservation in MHD supplemented by the induction equation. Additional source
terms include the pseudo-Newtonian gravitational potential of a BH and the
emission and absorption of neutrinos:
\begin{align}
  \label{eq:FLASHDensity} 
  \pder{\rho}{t} &+ \nabla\cdot{[\rho \v]} = 0 \\
  \label{eq:FLASHMomentum} 
  \pder{(\rho \v)}{t} &+ \nabla\cdot{[\rho(\v \otimes \v)-(\B \otimes \B)]}  + \nabla P = -\rho\nabla\Phi_{\rm A} \\ 
  \label{eq:FLASHEnergy}
  \pder{(\rho E)}{t} &+ \nabla\cdot{[\v( \rho E + P)-\B(\v\cdot\B)] } = -\rho\v\cdot \nabla \Phi_{\rm A} + Q_{\rm net}\\
  \label{eq:FLASHLepton}
  \frac{\partial Y_e}{\partial t} &+ \mathbf{v}\cdot\nabla Y_e = \Gamma_{\rm net}\\
  \label{eq:Induction}
  \pder{\B}{t} &+ \nabla\cdot{(\v\otimes\B - \B\otimes\v)} = 0,  
\end{align} 
where $\rho$ is the density, $\v$ is the velocity, $\mathbf{B}$ is the magnetic field 
(including a normalization factor $\sqrt{4\pi}$) , $Y_e$ is the electron fraction,
$E$ is the total specific energy of the fluid
\begin{equation}
E = \frac{1}{2}\left(\mathbf{v}\cdot\mathbf{v} + \mathbf{B}\cdot\mathbf{B}\right) + e_{\rm int},
\end{equation}
with $e_{\rm int}$ the specific internal energy, 
and $P$ is the sum of gas and magnetic pressure 
\begin{align} 
  P &= P_{\rm gas} + P_{\rm mag}, \\ P_{\rm mag} &= \frac{1}{2} \B \cdot \B.
\end{align} 

The induction equation (\ref{eq:Induction}) is discretized using the Constrained
Transport (CT) method \citep{Evans1988}
and conserved quantities are evolved using the
HLLD Riemann solver \citep{Miyoshi2005} with a piecewise linear
MUSCL-Hancock reconstruction method \citep{ColellaPLM1985}. 
The gravity of the BH is modeled with the pseudo-Newtonian
potential $\Phi_{\rm A}$ of \citet{Artemova1996}, ignoring the self-gravity of
the disk (see also \citealt{Fernandez2015a}).  The equation of state (EOS) is
that of \citet{TimmesSwesty2000}, with abundances of 
protons, neutrons and $\alpha$-particles in nuclear statistical equilibrium
(NSE) so that $P_{\rm gas} = P_{\rm gas}(\rho,e_{\rm int},Y_e)$, and accounting
for the nuclear binding energy of $\alpha$-particles as in \citet{FM13}.

We implement the framework for neutrino leakage emission 
and annular light bulb absorption described in \citet{FM13} and \citet{MF14}.
The scheme includes emission and absorption of electron neutrinos and antineutrinos
due to charged-current weak interactions
on nucleons, and with improvements in the calculation of the neutrino
diffusion timescale in high-density regions following 
the prescription of \citet{ILEAS2019}. A detailed description
of the implementation and verification tests (comparing to the
Monte Carlo scheme of \citealt{Richers2015}) are presented
in Appendix~\ref{AppB}.

The leakage and absorption scheme outputs scalar source terms for the
net rate of change of energy per unit volume $Q_\mathrm{net}$,
and net rate of change per baryon of lepton number $\Gamma_\mathrm{net}$,
which are respectively applied to $E$ and $Y_e$ (equations~\ref{eq:FLASHEnergy}
and \ref{eq:FLASHLepton}) in operator-split way.
We neglect the contribution of neutrinos to the momentum equation.

Finite-volume 
codes fail when densities in the simulation become too low. We impose a 
radial- and time-dependent density floor, designed to prevent 
unreasonably low simulation timesteps in highly magnetized regions 
(e.g., near the inner radial boundary and extending out a few km along the 
the rotation axis)
while also not affecting the dynamics of outflow. It has a functional form
approximately following that in \citet{Fernandez2019}
\begin{align}
\label{eq:density_floor} 
  \rho_\mathrm{floor} = \rho_\mathrm{sml}\l \frac{r}{20 \mathrm{km}} \r^{-3} \l
\frac{\mathrm{max}[t,0.1s]}{0.1 \mathrm{ s}} \r^{-1.5}, 
\end{align} 
where $\rho_\mathrm{sml} = 2\times 10^4 \text{ g cm}^{-3}$ and $r$ is the
spherical radius.
The time dependence is modelled after
empirically determining the rate of change of the maximum torus density in 2D
runs of the baseline model. When the density undershoots the floor value, it is topped
up to the floor level with material tagged as ambient, such that we can keep track of
it and discard it when assessing outflows and accretion. Keeping the density above the floor is
generally enough to prevent the internal energy (and gas pressure)
at levels that do not crash the code. Nevertheless, we also impose explicit floors for
these quantities, following the same form as in equation~(\ref{eq:density_floor}), 
but with normalizations $P_\mathrm{sml} = 2\times 10^{14}\,\text{erg\,cm}^{-3}$ and 
$e_\mathrm{sml} = 2\times 10^{11}$\,erg\,g$^{-1}$ for gas pressure and internal energy, respectively.

\subsection{Computational Domain and Initial Conditions} 

Equations~(\ref{eq:FLASHDensity})-(\ref{eq:Induction}) are solved in spherical 
polar coordinates $(r,\theta,\phi)$ centered at the BH and with the $z$-axis
aligned with the disk and BH angular momentum. The computational domain extends
from an inner radius, $r_\mathrm{in}$, located halfway between the innermost
stable circular orbit (ISCO) and the
BH horizon, both dependent on the BH mass and spin, to an outermost radius $r_\mathrm{out}$ located at
$10^4r_\mathrm{in}$.
The polar and azimuthal angular ranges are  $[5^\circ,175^\circ]$ and
$[0,180^\circ]$, respectively, corresponding to a half-sphere with a $5^\circ$ cutout around the
$z$-axis. The radial grid is discretized with $512$ logarithmically-spaced
cells satisfying $\Delta r/r \sim 0.018$, the meridional grid has $128$ cells
equally spaced in $\cos\theta$,  corresponding to $\Delta \theta \sim 0.92^\circ$ 
at the equator, and the azimuthal grid is uniformly discretized with $64$ cells.

The boundary conditions are set to outflow at the polar
cutout and at both radial limits, and to periodic at the $\phi$ boundaries.
The cutout around the polar axis is used to mitigate the stringent time step constraints
arising from the small size of $\phi$ cells next to the $z$ axis. 
We do not expect our polar boundary conditions to affect our analysis, as
the sub-relativistic outflow is well separated from the jet by a centrifugal
barrier \citep{Hawley2006}.
Any outflow along
the polar axes without the use of full GR is unreliable anyway, as many of
the proposed mechanisms for jet formation involve general relativistic energy
extraction from the BH spin energy (e.g., the Lense-Thirring and
Blandford-Znajek effects: \citealt{Bardeen1975,Blandford1977}). These processes
also involve the formation of a baryon-free funnel along the rotation axis, which 
means outflow along the polar axes contains minimal mass. Evolution tests using
reflecting, transmitting (with no azimuthal symmetry), and outflow polar
boundary conditions showed little to no difference over short times after initialization
($\sim$0.5 orbits). 

The initial condition for all of our models is an equilibrium torus 
with constant specific angular momentum, entropy, and composition,
consistent with the pseudo-Newtonian potential for the BH \citep{FM13,Fernandez2015a}. 
The input parameters are the BH mass, torus mass, radius of density peak,
entropy (i.e. thermal content or vertical extent), and $Y_e$.
In all cases, the latter two parameters are set to $s_B = 8 \;k_B/\text{baryon}$ 
and $Y_e = 0.1$, respectively, with other parameters changing between models
(\S\ref{s:models}).
Initial maximum tori densities $\rho_{\rm max}$ are typically 
$\sim 10^{10}-10^{11}$\,g\,cm${}^{-3}$.

Recent studies have shown that tori formed in NS mergers have a doubly peaked
distribution of $s_B$ and $Y_e$ \citep{Vsevolod2020,Most2021}, however, the use
of more realistic initial conditions for these quantities has little impact on
the resulting outflows (e.g., \citealt{Fujibayashi2020}), and is expected to be
smaller than differences due to our approximate handling of neutrino
interactions and gravity.

Models that start with a poloidal field are initialized
from an azimuthal magnetic vector potential which traces the density contours,
such that $A_\phi \propto \mathrm{max}(\rho-\rho_0,0)$, where $\rho_0$ is
defined as $0.009\rho_\mathrm{max}$, ensuring the field is embedded well within
the torus (e.g., \citealt{Hawley2000}).  This yields an initially poloidal field topology,
commonly known as ``standard and normal evolution" (SANE) in the literature.
The normalization is chosen such that 
the maximum magnetic field strength ($\sim 4\times10^{14}$\,G) is 
dynamically unimportant, with an average gas to
magnetic pressure ratio of 
\begin{equation} 
  \langle \beta \rangle = \frac{\int P_\mathrm{gas} \mathrm{d}V}{\int P_\mathrm{mag}\mathrm{d}V} = 100 
\end{equation}
with $\mathrm{min}(\beta)\sim 5$ at the inner edges and $\mathrm{max}(\beta)
\sim 10^5 $ at the initial density maximum. 
We also evolve a model that starts with a toroidal field, which is
initialized by imposing a constant 
$B_\phi =4\times10^{14}\,\mathrm{G}$
wherever $\rho > \rho_0$.
The magnetic field strength and mass density 
set the Alfv\'en velocities in the meridional and azimuthal directions,
\begin{equation}
  v^a_{\theta,\phi} = \frac{B_{\theta,\phi}}{\sqrt{\rho}}, 
\end{equation}
which in turn determine
the respective wavelengths of the most unstable MRI modes
(e.g., \citealt{Balbus1992, Duez2006}),
\begin{equation} 
  \lambda^\mathrm{MRI}_{\theta,\phi} \sim \frac{2\pi|v^a_{\theta,\phi}|}{\Omega_z}, 
\end{equation}
where $\Omega_z$ is the cylindrical angular velocity.  All of our simulations
resolve the relevant MRI modes with least 10 cells within the torus.
Resolution tests with 2D models indicate that our mass ejection results
have an uncertainty of $\sim 10\%$ due to spatial resolution (\S\ref{sec:resolution}).

\subsection{Models} 
\label{s:models}

\begin{table}
\caption{List of simulation parameters.
Columns from left to right show model name, BH mass, initial torus mass, initial
radius of maximum torus density, initial magnetic field geometry (pol: poloidal,
tor: toroidal), inclusion of the nuclear recombination energy of $\alpha$
particles in the EOS, use of neutrino absorption to evolve $E$ and $Y_e$, and
simulation dimensionality.  All BHs are assumed to have a dimensionless spin
parameter $0.8$.  }
\label{tab:ModelParams} 
\resizebox{\columnwidth}{!}{%
\begin{tabular}{lccccccc} 
\hline
Model & $M_\mathrm{bh}$ & $M_{\rm t}$ & $R_\mathrm{t}$ & $\mathbf{B}$ &
$\alpha$-rec & $\nu$-abs & dim \\ 
      & ($M_\odot$)     & $(M_\odot)$ & (km)           & geom & & & \\
\hline
\hline
base     & 2.65   & 0.10 & 50 & pol  & yes   & yes & 3   \\ 
bhns     & 8.00   & 0.03 & 60 &      &       &     &     \\ 
base-tor & 2.65   & 0.10 & 50 & tor  &       &     &     \\ 
\hline
base-2D    & 2.65   & 0.10 & 50 & pol  & yes   & yes & 2.5 \\ 
base-norec &        &      &    & pol  & no    &     &    \\ 
base-noirr &        &      &    &      & yes   & no  &    \\ 
\hline
\end{tabular}%
}
\begin{flushleft}
\end{flushleft}
\end{table} 

Table~\ref{tab:ModelParams} shows all the models we evolve and the parameters
used. Our \emph{base} model employs the most likely BH mass ($M_{\rm bh} = 2.65M_\odot$, dimensionless spin $0.8$), 
initial torus mass ($M_{\rm t} = 0.1M_\odot$),  and initial radius of density peak 
($R_{\rm t} = 50$\,km) for GW170817 (e.g., \citealt{LVSC2017a, Shibata2017,
FahlmanFernandez2018}), using a poloidal field geometry. Model \emph{bhns} uses 
a typical parameter combination expected from a BH-NS merger 
($M_{\rm bh}=8M_\odot$ with spin $0.8$; $M_{\rm t} = 0.03M_\odot$; $R_{\rm t}= 60$\,km), also 
with a poloidal initial field, to probe the effect of a higher
disk compactness (e.g., \citealt{Fernandez2020BHNS}).

Three additional simulations test the influence of key physical effects on the
\emph{base} model. Model \emph{base-tor} employs an initial toroidal magnetic
field geometry instead of poloidal. The other two are explored in axisymmetry
(2.5 dimensions): Model \emph{base-norec} sets the nuclear binding energy of
$\alpha$-particles to zero, and model \emph{base-noirr} turns off neutrino
absorption.  These two simulations are compared to model \emph{base-2D}, an
axisymmetric version of the \emph{base} model.  All 3D models are evolved for at
least $3$\,s, or until a time at which there is a clear power-law decay with
time in the ejected mass at large radius, allowing for an analytic
extrapolation until completion of mass ejection
(\S\ref{sec:outflow_characterization}). To achieve this phase, the \emph{base}
model needs to be evolved to $4$\,s.  The axisymmetric models are evolved until
$1.4\,$s, when accretion onto the BH stops due to a build up of magnetic
pressure: continuing evolution causes feedback which disrupts the torus. The MRI
is expected to dissipate in axisymmetry after $\sim 100$ orbits at the initial
torus density peak, corresponding to a few $100$\,ms (e.g.,
\citealt{Cowling1933,Shibata2007}).

\subsection{Outflow Characterization}
\label{sec:outflow_characterization}
The mass flux at a given radius is computed as
\begin{align}
  \dot{M}(r) =  \iint_{A_r} (\rho v_r  \d A_r),
\end{align} 
where the spherical area $A_r$ is given by
\begin{align} 
  A_r = \iint r^2 \sin\theta \d \theta \d \phi.  
\end{align} 
For outflows, the extraction radius is $r=10^4$ km, 
whereas for accretion onto the BH we take 
the radius of the ISCO.
We only consider unbound outflows, 
which we quantify with a positive Bernoulli parameter at the extraction radius
\begin{align} 
  \Phi_g + e_\mathrm{int} + e_\mathrm{k} + e_\mathrm{mag} +\frac{P_\mathrm{gas}}{\rho} > 0.
\end{align}
We also 
require that both outflowing and accreting matter have an atmospheric mass fraction
$\chi_\mathrm{atms} < 0.2$, and subtract off any remaining atmospheric mass so
that 
\begin{align} 
  \rho = \rho_\mathrm{tot}(1 - \chi_\mathrm{atms}).
\end{align} 
The total ejected mass is computed by temporally integrating the outflow
mass flux, such that
\begin{align}
\label{eq:Mtot_simulation}
  M_\mathrm{out} =  \int_t \iint_{A_r} (\rho v_r  \d A_r) \d t.  
\end{align} 

\begin{table*} 
\caption{
Mass ejection results for all our simulations. Columns from left to right show
model name, total unbound mass ejected during the first $3\,$s of the
simulations for 3D models and first $1.4\,$s for 2D models
(equation~\ref{eq:Mtot_simulation}, in percent of a solar mass and relative to
the initial torus mass $M_{\rm t}$), $\dot{M}_{\rm out}$-weighted average electron
fraction (equation~\ref{eq:Ye_ave}) and radial velocity
(equation~\ref{eq:vr_ave}), maximum simulation time, as well as unbound ejected
mass, average electron fraction, and radial velocity broken down by electron
fraction (superscript \emph{blue} -- lanthanide-poor: $Y_e \ge 0.25$, \emph{red}
-- lanthanide-rich: $Y_e < 0.25$).  The last column shows the range of total
unbound ejected mass extrapolated to infinity with a power-law fit to the mass
outflow rate (equation~\ref{eq:Mtot_extrapolated}). Note that 2D and 3D
simulations run for different times, so direct comparisons of simulations
with different dimensionality should not be made (See
\S\ref{sec:dimensionality}).} 
\centering
\resizebox{\textwidth}{!}{%
\begin{tabular}{l|cccccccccccc} 
Model
& \multicolumn{2}{c}{$M_\mathrm{out}$} & $ \la Y_e \ra              $ & $\la v_r               \ra $ & $t_{\rm max}$
& $M^\mathrm{blue}_\mathrm{out} $ & $ \la Y_e^\mathrm{blue}\ra $ & $\la v^\mathrm{blue}_r \ra $  
& $M^\mathrm{red}_\mathrm{out}  $ & $ \la Y_e^\mathrm{red} \ra $ & $\la v^\mathrm{red}_r  \ra $ 
& $M_\mathrm{out}^\mathrm{extr} $ \\
& $ (10^{-2} M_\odot)           $ & ($M_{\rm t}$) &              & $ (c) $ & (s)                      
& $ (10^{-2} M_\odot)           $ &                              & $ (c) $                       
& $ (10^{-2} M_\odot)           $ &                              & $ (c) $ 
& $ (10^{-2} M_\odot)           $ \\ 
\hline\hline

base      & 2.750 & 0.275 & 0.195 & 0.105 & 4.1 & 0.660 & 0.302 & 0.087 & 2.090 & 0.161 & 0.110 & 3.55-4.44 \\
bhns      & 0.155 & 0.052 & 0.197 & 0.065 & 4.0 & 0.014 & 0.260 & 0.057 & 0.140 & 0.190 & 0.064 & 0.18-0.23 \\
base-tor  & 4.237 & 0.424 & 0.173 & 0.079 & 3.0 & 0.655 & 0.291 & 0.100 & 3.583 & 0.152 & 0.075 & 4.76-7.01 \\
\noalign{\smallskip}
\hline
base-2D   & 6.225 & 0.623 & 0.329 & 0.182 & 1.4 & 5.528 & 0.349 & 0.169 & 0.697 & 0.176 & 0.279 & -         \\
base-norec  & 6.146 & 0.615 & 0.311 & 0.158 & 1.4 & 5.275 & 0.330 & 0.151 & 0.870 & 0.195 & 0.199 & -         \\
base-noirr  & 3.323 & 0.332 & 0.205 & 0.229 & 1.4 & 0.996 & 0.298 & 0.321 & 2.327 & 0.165 & 0.190 & -         \\
\hline
\end{tabular}%
}
\label{tab:ModelResults} \end{table*}

The mass weighted averages of electron fraction and radial velocity,
\begin{align} 
\label{eq:Ye_ave}
  \la Y_e \ra = \dfrac{\int_t \iint_{A_r} (\rho v_r Y_e \d A_r) \d t, }{M_\mathrm{out}} \\ 
\label{eq:vr_ave}
  \la v_r \ra = \dfrac{\int_t \iint_{A_r} (\rho v_r v_r \d A_r) \d t, }{M_\mathrm{out}}
\end{align} 
are provided as a summary of our model results in
Table~\ref{tab:ModelResults}.  We further subdivide outflows based on their
electron fraction into ``red" ($Y_e < 0.25$) and ``blue" ($Y_e \geq 0.25$), 
based on kilonova models which predict a sharp
cutoff between lanthanide-rich and lanthanide-poor matter
(e.g., \citealt{Kasen2015,Lippuner2015}).

Once the torus reaches a quasi-steady phase following freezeout of weak
interactions, ($t_{ss} \sim 1.1\mathrm{s}$), the mass outflow rate enters a
phase of power-law decay, $ \dot M_\mathrm{out}(t > t_\mathrm{ss}) \propto
t^{-\delta}$.  We can therefore estimate the completed mass ejection over
timescales of $\sim 10$\,s by extrapolating from a power-law fit to the mass outflow rate (e.g.,
\citealt{Margalit2016, Fernandez2019WD}),
\begin{equation}
\label{eq:Mtot_extrapolated}
  M_\mathrm{out}^\mathrm{extr}  = M_\mathrm{out}(t_{\rm ss}) + \frac{1}{\delta -1} \dot
M_\mathrm{out}(t_\mathrm{ss}) t_\mathrm{ss},
\end{equation}
where the integral in equation~\ref{eq:Mtot_simulation} is computed until $t=t_{\rm ss}$.
The choice of $t_{\rm ss}$ is made based on visual inspection of when the
cumulative mass outflows begin to plateau.  Varying this choice in response to
episodic mass ejection events results in an uncertainty in the exponent of 
$|\Delta \delta| \lesssim 0.5$ corresponding to about 5-15\% difference in total
$M_\mathrm{out}^\mathrm{extr}$. 

\section{Results} 

\label{sec:results} 
\begin{figure*} 
\includegraphics[width=\textwidth]{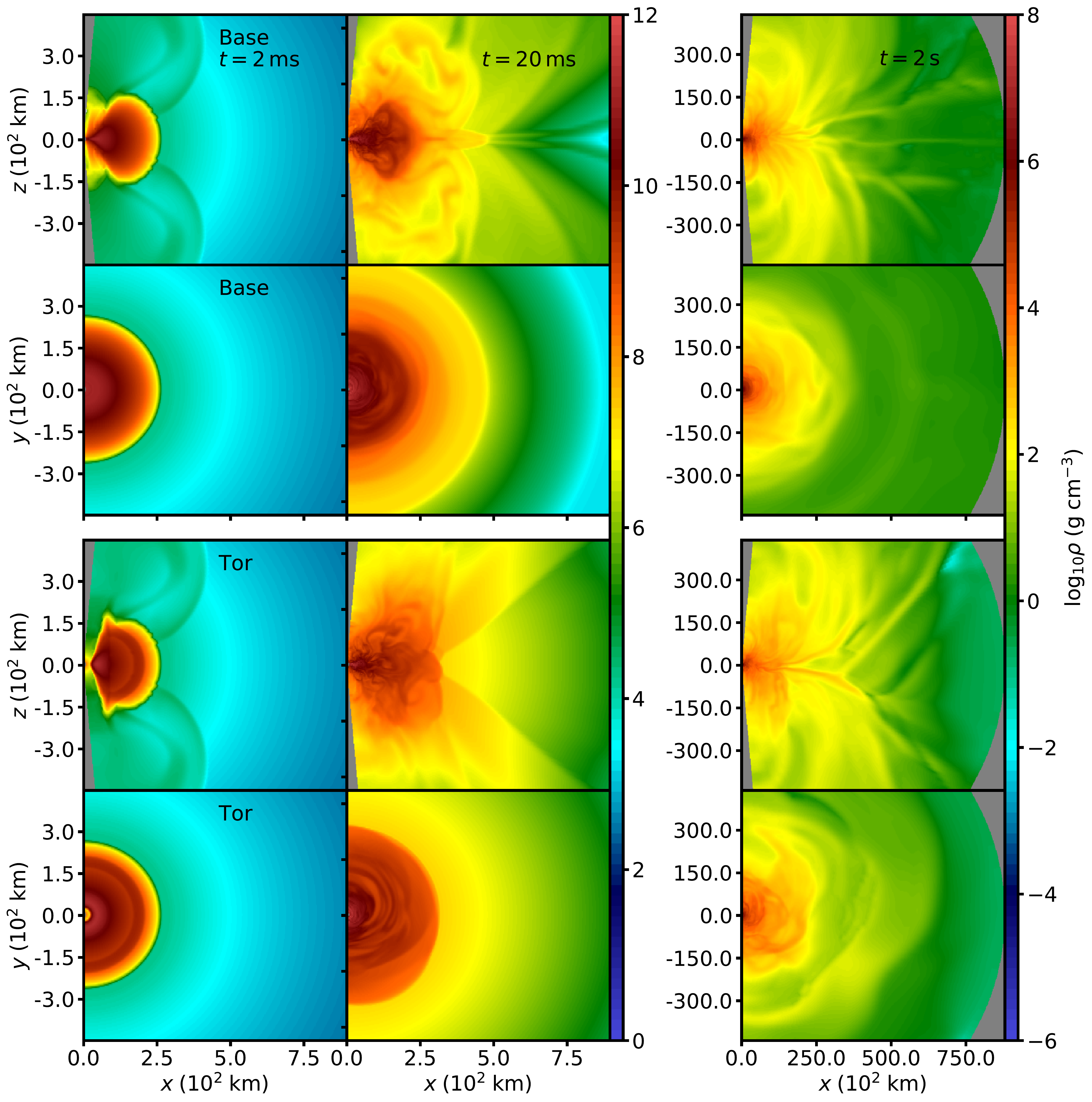}
\caption{Density slices in the  $x-z$ ($\phi=0$) and $x-y$ ($\theta=0$) plane of the
\emph{base} (top two rows) and \emph{base-tor} (bottom two rows) models at t=2\,ms (0.6 orbits at the initial torus density peak), 20\,ms (6
orbits), and 2\,s (600 orbits).  The first two columns correspond to the early
accretion and initial wind phase. Note the change in both colour table and length
scale for the final column. Gray regions are outside the computational domain.}
\label{fig:BaseTorus2DPlots}
\end{figure*} 

\subsection{Overview of Torus Evolution in MHD}

Our \emph{base} and \emph{bhns} runs, with a poloidal field embedded in the torus, 
show very similar evolution to previous runs in (GR)MHD. 
Within a few orbits, accretion onto the central object begins as magnetic
pressure in the torus builds up via winding and onset of the MRI, disrupting
hydrostatic equilibrium. In the \emph{base-tor} run, accretion onto the
torus begins as turbulence driven within the torus by the toroidal MRI generates
poloidal field. Both runs then begin mass ejection as the poloidal MRI
grows (Figure~\ref{fig:BaseTorus2DPlots}). 

In contrast to hydrodynamic models, mass ejection begins on a timescale of
$\sim$ms, forming ``wings" of ejected material away from the midplane and
rotation axis. More isotropic, thermally-driven ejecta takes over at $\sim$1\,s,
as neutrino emission has subsided and the disk enters an advective state. As
material moves outward, it cools and releases the binding energy stored in
$\alpha$-particles, increasing the internal energy of the fluid.  Neutrino
absorption, although subdominant energetically, is important in driving the
evolution of $Y_e$. By $\sim$1\,s, the torus has reached an equilibrium value of
$Y_e$, and the cumulative mass outflow begins to plateau. 


\subsection{Mass Ejection in 3D Models}

Table~\ref{tab:ModelResults} shows the total unbound mass ejected by the end of
each simulation (equation~\ref{eq:Mtot_simulation}), and the extrapolation of of
the mass outflow rate to infinity in time (equation~\ref{eq:Mtot_extrapolated}),
for all of our 3D models.  The \emph{base} and \emph{base-tor} models eject
$\sim 28\%$ and $\sim 42\%$ of the initial torus mass during the simulation,
respectively, with the extrapolated mass ejected at late times being $\sim 40\%$
and $\gtrsim 50\%$.  The \emph{bhns} model ejects the least mass owing to the
high compactness, with $\sim 5\%$ of the initial torus mass ejected by the end
of the simulation.  All 3D simulations show  average velocities in the range
$\langle v_r \rangle \sim (0.07-0.11)\,c$, and average electron fractions
$\langle Y_e \rangle \sim 0.17-0.20$. 

\begin{figure}
\includegraphics[width=\columnwidth]{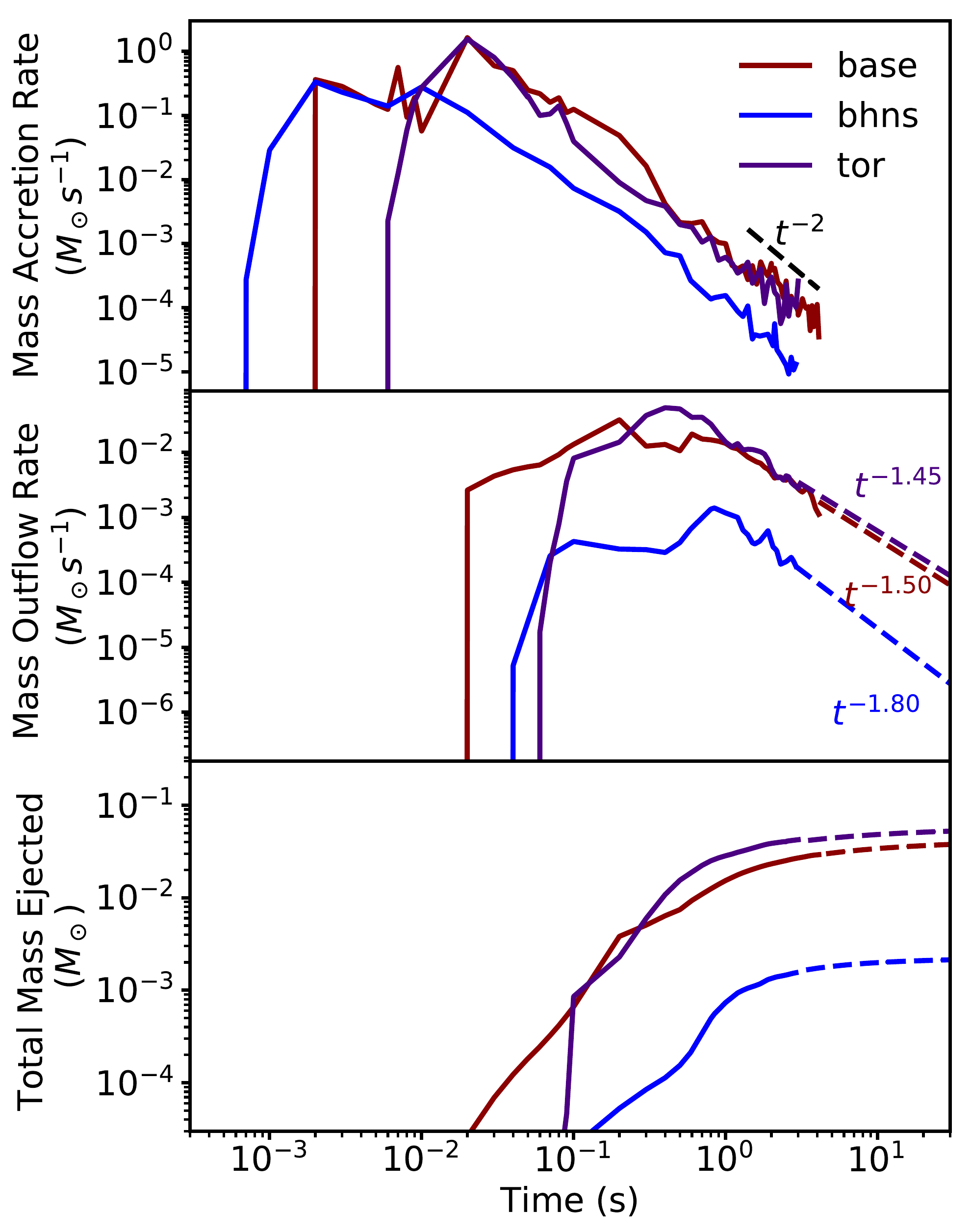}
\caption{\emph{Top:} mass accretion rate at the 
ISCO for all 3D models, as labeled. The dashed line shows a reference
power-law dependence $t^{-2}$. 
\emph{Middle:}
unbound mass outflow rate at an extraction radius $r=10^4$\,km, for all
3D models. Outflow rates are extrapolated with a power law (dashed lines), 
with the fiducial power law index shown for each curve (equation~\ref{eq:Mtot_extrapolated}).
\emph{Bottom:}
evolution of the cumulative unbound mass ejected (equation~\ref{eq:Mtot_simulation}). 
Dashed lines show the result of using the same power-law extrapolation of the outflow
rate after the end of the simulation (equation~\ref{eq:Mtot_extrapolated}).}
\label{fig:MassAccretionOutflows}
\end{figure}

\begin{figure}
\includegraphics[width=\columnwidth]{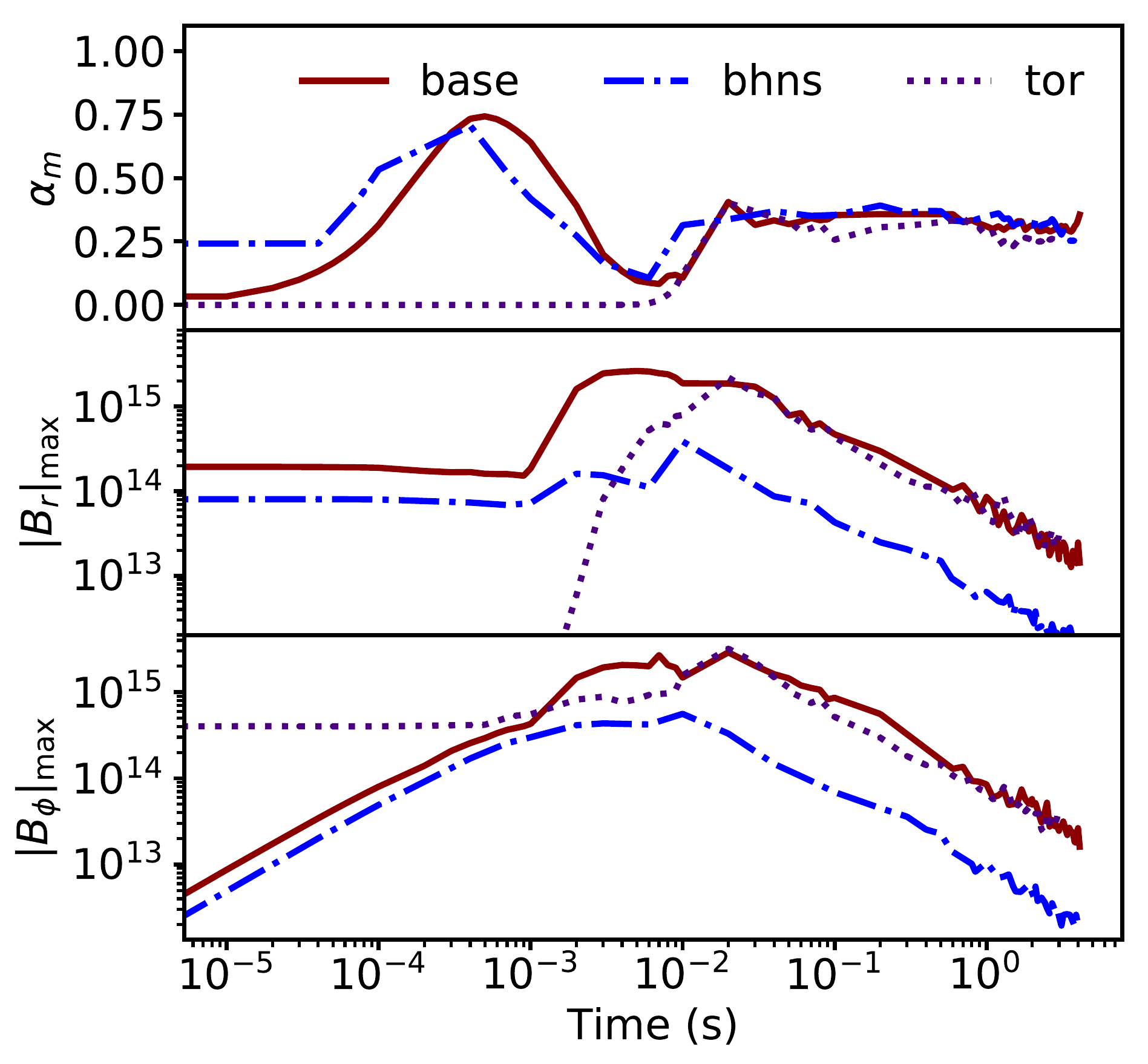}
\caption{\textit{Top}: 
Evolution of the volume-averaged Maxwell stress normalized by the volume-averaged magnetic pressure, 
$\alpha_m=\langle B_r B_\phi \rangle/\langle P_\mathrm{mag}\rangle$,
a measure of radial angular momentum
transport by magnetic forces (see \citealt{Hawley2000,Hawley2001} for a
discussion). \textit{Middle}: Evolution of the maximal magnitude of the radial magnetic field.
\textit{Bottom}: Evolution of the maximal magnitude of the azimuthal magnetic field. 
}
\label{fig:MagneticGrowth}
\end{figure}

The mass accretion and outflow rates for all 3D models are shown in 
Figure~\ref{fig:MassAccretionOutflows}.  Each model begins to eject mass at a
steeply rising rate, primarily due to MHD effects, which eventually 
reaches a plateau. After
rising to a peak at time $\sim 1\,$s, the mass ejection rate then begins to
decay as a power law with episodic ejection events. Figure~\ref{fig:MassAccretionOutflows}
also shows that by this time, the cumulative mass ejected begins to plateau.
Differences between models
manifest as changes in the initial outflow time: the \emph{base-tor} and
\emph{bhns} runs begin to eject matter $\sim 0.04\,$s and $0.02\,$s later than the
\emph{base} model, respectively. In the \emph{base-tor} case, this delay is
caused by the additional time required for the toroidal MRI to generate poloidal
field, which then drives angular momentum transport. This is illustrated in
Figure~\ref{fig:MagneticGrowth}, which shows the evolution of the
volume-integrated Maxwell stress for all 3D models.  The \emph{bhns} run has a
deeper gravitational potential at the initial density maximum than the other
runs (see \S\ref{sec:Compactness}), leading to more total energy input required
to begin mass ejection (e.g., \citealt{Fernandez2020BHNS}). 

We also find that mass ejection peaks earlier in the \emph{base} model than in
the other two runs. The \emph{base-tor} run reaches peak mass ejection around
0.1\,s later than the base model at a somewhat larger outflow rate, but then
decays with time following a power law slope of $\delta_\mathrm{tor} = 1.45$,
only $\sim$3\% different than the \emph{base} model slope $\delta_\mathrm{base}
= 1.50$. This qualitatively similar behaviour between poloidal and toroidal
models is also found by \citet{Christie2019}, although the initial conditions of
their simulations lead to different quantitative values of $\delta$.
Each 3D model has a different quantitative value of $\delta$,
despite having the same late-time mass ejection mechanism. This variation can be
attributed to physical differences in the disks and the timing of mass ejection.
Each disk reaches maximum outflow at a phase in its evolution when the 
remaining disk mass, wind loss rate, and accretion rate are different relative
to the initial disk mass and timescale of angular momentum
transport in the disk. The initial time of mass ejection is also related to the range of electron
fractions in the outflow. All runs produce a broad range of $Y_e$ in the ejecta,
with a lower limit $Y_e \gtrsim 0.05$ 
(See \S\ref{sec:Neutrinos}). 

\subsubsection{Morphology}
\label{sec:Morphology} 

\begin{figure*}
\centering \includegraphics[width=0.8\textwidth]{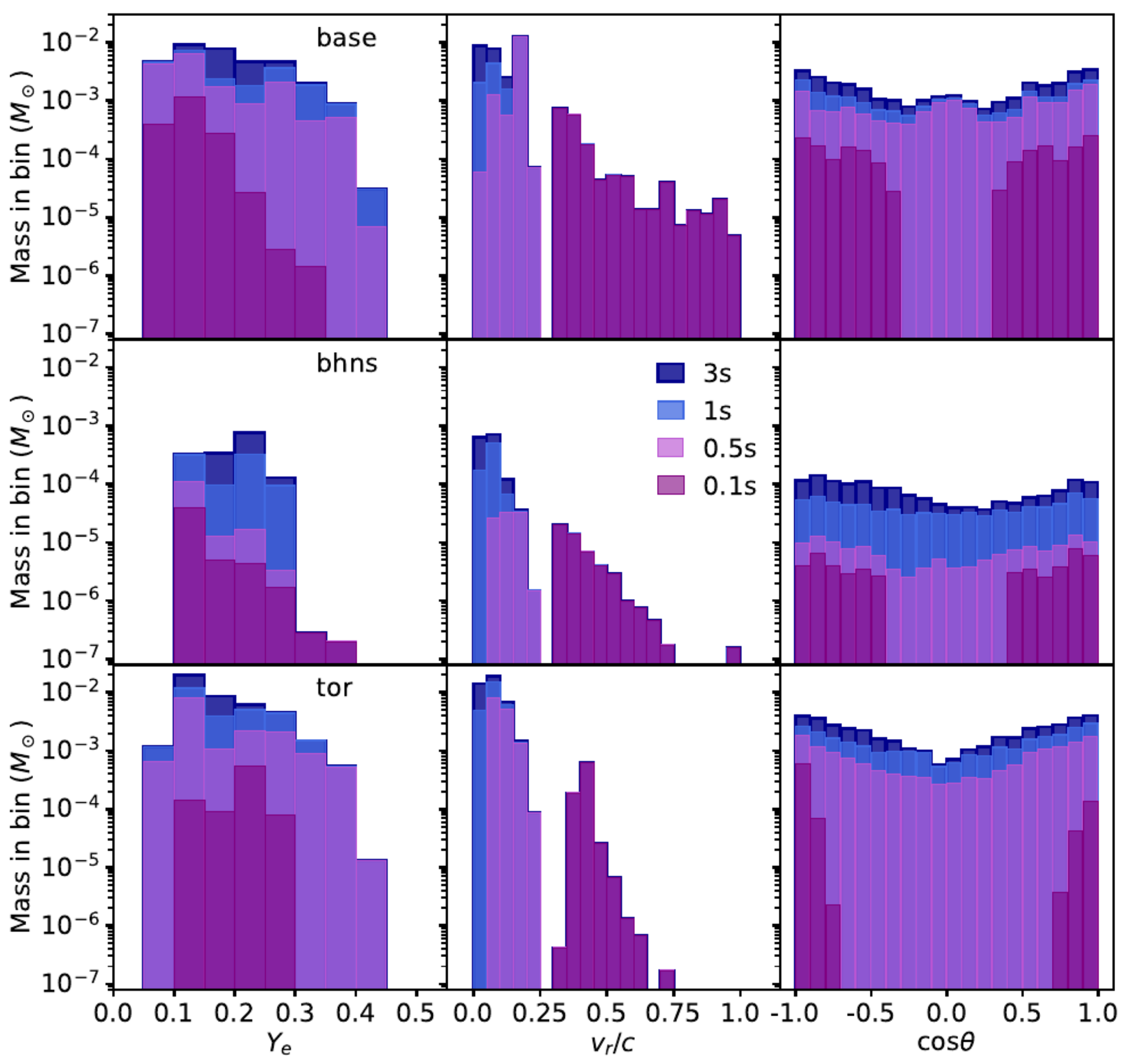} 
\caption{Total unbound mass ejected by our 3D models, as labeled, sorted into bins based on electron fraction $Y_e$,radial velocity $v_r$, and polar angle $\theta$. The bin sizes are $\Delta Y_e =
0.05$, $\Delta v_r/c = 0.05$, and $\Delta \cos\theta = 0.1$, where $\cos\theta =
0$ is the midplane.} 
\label{fig:Histograms} 
\end{figure*}

Kilonova emission is dependent on the ejecta morphology (e.g.,
\citealt{Kasen2017, Kawaguchi2019, KawaguchiBNS2020,Korobkin2021,Heinzel2021}),
which can vary depending on the type of binary and mass ejection mechanism.
The morphology of the disk outflow ejecta for the \emph{base} and
\emph{base-tor} models is shown in the rightmost panel of
{Figure~}\ref{fig:BaseTorus2DPlots}, showing a characteristic ``hourglass" shape
found in previous GRMHD simulations (e.g.,
\citealt{Fernandez2019,Christie2019}). This feature is robust across all our 3D
models, as can be seen from the angular mass outflow histograms in
Figure~\ref{fig:Histograms}.  Model \emph{base-tor} ejects $50\%$ less mass with
$v\gtrsim 0.25\,c$ and within $\sim \pi/4$ of the rotation axis than model
\emph{base}, and no ejecta in this velocity range is produced within $\sim
\pi/4$ of the equatorial plane.  In contrast, models \texttt{base} and
\texttt{bhns} have a much wider distribution of fast/early ejecta, extending
down to within $\pi/6$ of the rotation axis.  This implies that the morphology
of the highest velocity ejecta is dependent on the initial condition of the
torus, with the compactness of the disk having little effect on outflow
geometry.  This result is supported by the results of \citep{Christie2019}, as
well as those of \citep{Siegel2018} in their discussion of the initial transient
phase.

\subsubsection{Compactness}
\label{sec:Compactness} 

We find that the fraction of the initial disk mass ejected decreases with
increasing disk compactness (model \emph{bhns}), following the same trend as the
hydrodynamic results of \citet{Fernandez2020BHNS}. This shows that the
additional mass ejected by MHD effects relative to pure viscous hydrodynamics
also decreases with increasing disk compactness. Figure~\ref{fig:Histograms}
shows that the fastest ejecta ($t<0.1s$) becomes a smaller fraction of the total
mass ejected, with the majority of the disk outflow ensuing after $1$\,s.  This
change can be attributed to multiple differences with the \emph{base} model: the
gravitational potential is deeper by a factor $\sim 2$ at the density maximum,
the ISCO of the BH is closer to the torus density maximum (see, e.g.,
Figure~\ref{fig:MassAccretionOutflows}), and the initially lower density torus
emits an order of magnitude less neutrino luminosity and is more transparent to
neutrinos. Nevertheless, the different disk structure results in a shorter
neutrino cooling time at the initial density peak in the \emph{bhns} model ($1$\,ms)
relative to the \emph{base} model ($2$\,ms). Thus, at early times ($t < 0.1s$) when neutrino
cooling is strong, disk material is more bound in the
\emph{bhns} than in the \emph{base} case.

We note a sharp  drop in mass ejected with $Y_e\sim0.3$ for the \emph{base}
model, also found in \citet{Fernandez2020BHNS}, which can be attributed to the
larger relative importance of neutrino absorption in more massive tori.  For a
neutron-rich disk where neutrino absorption dominates the evolution of $Y_e$,
the process $\nu_e + n \rightarrow e^- + p$ occurs more frequently than its
inverse, increasing the net $Y_e$
\citep{Siegel2018,Fernandez2020BHNS,Most2021}.This trend is also found by
\citet{just_2021} with a significantly more advanced neutrino scheme - a broader
distribution in $Y_e$ corresponds to more absorption (see their Figure 13).
Similar 2D axisymmetric simulations by \citet{Shibata2012} yield neutrino
luminosities that decrease from $10^{52}$ erg s$^{-1}$ to $10^{51}$
erg\,s$^{-1}$ as the BH mass increases by from $3M_\odot$ to $6 M_\odot$. We
find a very similar trend as we change the compactness. 

\subsection{Mass Ejection in 2D: Sensitivity to Physics Inputs}

\subsubsection{Dimensionality}
\label{sec:dimensionality}
Measuring unbound ejecta by the end of
each simulation, the \emph{base-2D} model ejects a factor of $\sim 2$ more mass than the
equivalent \emph{base} run in 3D, despite running for $1.4\,$s instead of
$3\,$s. The evolution is qualitatively similar for the first $\sim1\,$s, until
the axisymmetric torus becomes dominated by magnetic pressure and is disrupted,
at which point we end the simulation. Quantitatively, model \emph{base-2D}
produces a higher mass outflow rate at all times, in particular more ejecta with
velocities $\gtrsim 0.25\,c$.  We attribute this enhanced mass ejection to the
lower accretion rate onto the BH in axisymmetry given the suppression of the
MRI, with divergence in the evolution from the 3D case starting at $\sim
20$\,ms.  With less accretion, the larger amount of matter in the torus results
in a higher outflow rate, given that the same outflow driving processes operate
in 2D and in 3D.

\subsubsection{Spatial Resolution}
\label{sec:resolution}
To quantify uncertainties due to spatial resolution, we run versions of
model \emph{base-2D} at half and twice the resolution in both the radial and
polar directions.  The high-resolution model is evolved until $0.5$\,s, probing
the early, magnetically-driven phase, and the half-resolution model is evolved
until $1$\,s, which includes the radiatively-inefficient phase of mass ejection.
Mass ejection up to $1$\,s is $\sim10$\% higher in the standard resolution model
relative to the low-resolution model. We thus associate an uncertainty of 10\%
to our mass ejection numbers due to spatial resolution. The growth of the
toroidal magnetic field is identical during both the magnetic winding and MRI
growth phase in all models until $3$\,ms, when the maximum value of toroidal
magnetic field saturates at $2\times 10^{15}$\,G. We find a 0.5\% difference in
the saturation value of $|B_\phi|$ between the standard and high resolution
runs, and a 5\% lower saturation value comparing the low to standard resolution
run.  Thereafter, the maximum of $|B_\phi|$ undergoes stochastic fluctuations
with amplitude of order unity until a dissipation phase begins at $\sim 30$\,ms.
The standard and high-resolution models remain consistent within fluctuations.

\subsubsection{Nuclear Recombination}
Comparing mass ejection from models \emph{base-2d} (with recombination) and
\emph{base-norec} (without recombination), we find
that nuclear recombination remains a subdominant effect until the end of our 2D
simulation at $1.4\,$s. Before $0.5\,$s, energy input from nuclear recombination
increases the mass-averaged velocity of ejecta, resulting in a noticeable
decrease in mass ejected at $\sim 0.3\,c$ and a increase at $\sim0.25\,c$ in
model \emph{base-norec} compared to model \emph{base}. After $\gtrsim 0.5\,$s,
comparatively less mass is ejected in model \emph{base-norec} due to the lack of
recombination heating. In other words, mass which would have been ejected in the
initial MHD-driven phase is instead ejected slower and at a later time,
indicating that the net effect of nuclear recombination is to make matter less
gravitationally bound and thus easier to eject by magnetic forces at early
times.  We find an almost identical distribution in electron fraction, skewed to
a slightly lower average value since less mass is ejected later when the charged
current weak interactions have already raised the $Y_e$.  

\subsubsection{Neutrino Absorption}
\label{sec:Neutrinos}
Inclusion of neutrino absorption results in additional mass ejection by a factor
of $\sim 2$ relative to a model without it (\emph{base-noirr}), and a negligible
effect on the mass ejected at $t <0.1$\,s, when magnetic stresses dominate.  The
energy input from neutrino absorption causes additional mass to become
marginally unbound, extending the distribution in velocity space to slower
outflow. However, since more mass is ejected, neutrino absorption produces a
\textit{decrease} by $0.05\,c$ in the mass averaged velocity. Turning off neutrino
absorption skews the electron fraction of the ejecta to lower values, with more
mass (factor of 2) being ejected at all times with $Y_e < 0.1$, and 2 orders of
magnitude less ejecta with $Y_e > 0.4$.

\subsection{Comparison to previous work} 

The ejected masses and velocities from our 3D models are in broad agreement with
comparable simulations \citep{Siegel2018, Miller2019,
Fernandez2019,Christie2019,just_2021}. The \emph{base} run is qualitatively
closest to the model of \citet{Siegel2018}, which lacks neutrino absorption,
and to the MHD model of \citet{just_2021}, which has a less massive torus ($0.01
M_\odot$). \citet{Siegel2018} find that 16\% of the torus is ejected during 381
ms of evolution, and \citet{just_2021} that 20\% of the initial torus mass is
ejected during 2.1\,s.  Our \emph{base} model ejects a higher fraction of the
initial disk mass due to the difference in compactness as well as a longer
simulation time.  Relative to the long-term GRMHD simulation of
\citet{Fernandez2019}, which employed an initial field geometry conducive to a
magnetically-arrested disk (MAD, e.g., \citealt{Tchekhovskoy2011}), ran for a
longer time ($\sim 10$\,s), and did not include neutrino absorption, our base
run ejects $\sim 20\%$ less mass by the end of the simulation at 3s. 
Our extrapolated ejected masses are comparable to that from this longer run,
with other differences explainable by the difference in compactness 
and initial field geometry.  The weak poloidal (SANE) model from
\citet{Christie2019} is run for the same amount of time as our base run but with
an initially weaker field, and also ejects $\sim 30\%$ of the torus mass during
the simulation, despite not including neutrino absorption.

The toroidal run of \citet{Christie2019} ejects 3\% less mass than their
weak poloidal (MAD) run, whereas we find 15\% more mass ejection in our \emph{base-tor} (toroidal)
model relative to our \emph{base} (poloidal) model.  We do find a lower average
velocity (by $\sim 0.02\,c$) in the \emph{base-tor} model relative to our
\emph{base} model, same as they do. 
\citet{Christie2019} find that their toroidal model begins mass ejection
at almost the same time as their weak poloidal run, and find a more sustained period
of mass ejection from $\sim 0.01-0.05$\,s in the toroidal model. Comparatively, we find that mass
ejection in our \emph{base-tor} model begins later and quickly rises to peak at
a value higher than that of the poloidal simulation.  This difference in
dynamics could be attributed to a comparatively stronger toroidal magnetic field
($\beta \sim 0.01-2$ in the initial torus) and the effect of neutrino absorption. 
We do not vary the initial field strength in our simulations, as a
lower field strength would require more cells to properly resolve the
MRI. We can speculate on how lowering the field strength would change our
outflows by comparing to the results of \citet{Christie2019}. They find
that lowering the field strength reduces the initial ($t \lesssim 0.5$\,s) outflows
driven by magnetic stresses, but the late-time thermal outflows are nearly identical.
The effects on our (SANE) field configuration would likely be similar, but less
prominent, given that their MAD configuration is optimized for producing magnetic outflows.

The work of \citet{Miller2019} utilizes the same initial conditions as our
\emph{base} model but with a more advanced neutrino scheme to treat neutrino
emission and absorption. We find a similar amount of mass ejected by 100ms of
evolution ($\sim 2\times 10^{-3} \msun$), indicating broad agreement despite the
difference in neutrino schemes.  

Utilizing a mean field dynamo to address the suppression of MRI in 2D,
\citet{Shibata2021MeanField} run resistive MHD simulations of high compactness
toroidal disks. They find $\sim10-20\%$ of the initial disk mass is ejected over
$\gtrsim 4\,$s with an average electron fraction $\langle Y_e \rangle \sim
0.25-0.35$, in broad agreement with our findings. Notably, they find that mass
ejection begins $\gtrsim 500$\, ms later than in our toroidal run, although this
delay can be attributed to the high compactness of their models.

The recent GRMHD simulations of \citet{Hayashi2021} 
start from the inspiral of a 1.35$\msun$ NS and a 5.4 or 8.1 $\msun$ BH and evolve
the remnant for up to $2$\,s, including neutrino leakage and
absorption. They find qualitatively similar results when compared to our runs,
albeit with much larger tori masses post-merger ($\sim 0.2-0.3 \msun$). The
fraction of the torus mass ejected is also comparable to our \emph{bhns} runs,
as $\sim 10\%$ of their tori is ejected in the first 1\,s, with a broad
distribution in both electron fraction and velocity. Discounting dynamical
ejecta, they find a peak electron fraction $ Y_e \sim 0.25-0.35$ and post merger
outflow velocities of $v \lesssim 0.08\,c$, in good agreement with our results.
They find outflows starting at $\gtrsim 200$\,ms later than our \emph{bhns}
model, however the initial tori in their simulations are in a deeper potential well, and form with
an initially toroidal field, both of which we find delay outflows in comparison
to our \emph{base} model.  

By analyzing the net specific energy of tracer particles, \citet{Siegel2018} find
that nuclear recombination of $\alpha$-particles plays a key role in unbinding
matter in the disk outflow (in a simulation that does not include neutrino 
absorption). Our 2D model \emph{base-norec} which has nuclear recombination turned
off but includes neutrino absorption, ejects only slightly less mass than our 
\emph{base-2D} run 
indicating that under these circumstances
nuclear recombination is a sub-dominant effect. It remains to be tested
whether recombination will remain sub-dominant in a fully 3D simulation that
includes neutrino absorption and runs for a long time ($\gg 1$\,s).

Our 2D model without neutrino absorption (\emph{base-noirr}) ejects a factor $\sim 2$
less mass than the \emph{base-2D} model. This difference is significantly larger
than that found in models that employ viscous hydrodynamics, which typically find
that neutrino absorption is dynamically sub-dominant for mass ejection (e.g.,
\citealt{FM13,Just2015}). This also inconsistent with the 3D MHD run of
\citet{just_2021}, who find that turning off neutrino absorption
results in a 2.5\% \textit{increase} in mass ejection relative to the initial
torus mass, although they use a
different neutrino leakage scheme and a less massive torus that is more transparent to
neutrinos.
Our results are limited by the use of axisymmetry for these simulations, but suggest that 
neutrino absorption could indeed be more significant for the dynamics
of mass ejection and motivates further studies in 3D.

\section{Summary and Discussion} 
\label{sec:conc}

We have run long-term 3D MHD simulations to explore mass ejection from BH-tori
systems formed in neutron star mergers. The publicly available code {\tt
FLASH4.5} has been extended to allow its unsplit MHD solver to work on
non-uniform spherical coordinates in 3D (Appendix~\ref{AppA}, \ref{AppB}). All
of our models include a physical EOS, neutrino emission and absorption via a
leakage scheme with disk-lightbulb irradiation (Appendix~\ref{AppC}), and treat
the gravity of the BH with a pseudo-Newtonian potential.  Our 3D models employ
different initial magnetic field geometries and disk compactnesses.  We have
also carried out axisymmetric models that suppress the
nuclear recombination and neutrino absorption source terms.

The disk outflows from our 3D models exhibit a broad distribution in electron
fraction and ejection polar angle (Figure~\ref{fig:Histograms}), with a typical
hourglass morphology (Figure~\ref{fig:BaseTorus2DPlots}). The tori eject matter
with a bimodal distribution in velocity (Figure~\ref{fig:Histograms}) associated
with two different mass ejection phases: MHD stresses power early time ($t <
0.1$\,s) high velocity ($v \gtrsim 0.25\,c$) ejecta, and late-time ($t\sim$\,s)
``thermal" ejection provides the majority of mass outflows centered around $v
sim 0.1\,c$.
 
We find that imposing an initially toroidal field configuration ejects
$\sim$15\% more of the initial torus mass than the standard SANE poloidal field
of similar maximal field strength (Figure~\ref{fig:MassAccretionOutflows},
Table~\ref{tab:ModelResults}).  However, the toroidal model ejects an order of
magnitude less ejecta in the first 100\,ms of evolution
(Figure~\ref{fig:Histograms}), comprising all of the ejecta travelling at
velocities $v \gtrsim 0.25\,c$. The high-velocity ejecta is suppressed by the
additional time it takes for dynamo action to convert toroidal into large-scale
poloidal fields, which then drives radial angular momentum transport.  

Increasing the disk compactness to values expected for typical BH-NS mergers
results in significantly
less mass ejection relative to our \emph{base} (NS-NS) model, beginning at a
later time and decaying at a faster rate.
Comparing to the viscous hydrodynamic models of \citet{Fernandez2020BHNS}, we
find the same overall trend of decreased mass ejection with increasing
disk compactness. Model \emph{bhns} has the same initial torus and BH configuration
as their model \emph{b08d03}: by 4\,s, our 3D MHD simulation has ejected $5.5\%$
of the initial torus mass, while the viscous hydrodynamic equivalent has ejected
only $1.9\%$ of the initial disk mass by the same time. The hydrodynamic model goes on to eject $5.0\%$
of the initial disk mass after $12\,$\,s of evolution, with mass ejection peaking at
$2.3$\,s but remaining non-negligible until later times. The extrapolated outflow
for our 3D MHD model \emph{bhns} predicts $6-8\%$ of the initial torus mass, which is
consistent with the previously-found enhancement in mass ejection by MHD relative to viscous
hydrodynamics at smaller
compactnesses when evolving both to $\sim 10$\,s \citep{Fernandez2019}.
Our results inform analytic fits to fractions of the initial 
disk mass ejected like that of \citet{raaijmakers_2021}, which
has the enhancement in mass ejection due to magnetic effects relative to viscous
hydrodynamics as a free parameter. More 3D MHD simulations at different
compactness and with various initial magnetic field geometries are needed
to improve  the predictive power of these fits.

Our axisymmetric models that vary the physics show that nuclear recombination is
a sub-dominant effect, while neutrino absorption can make a significant
difference in mass ejection.  Inclusion of neutrino absorption produces a shift
of the velocity distribution at late times, down by $\sim 0.05\,c$, with
negligible effects at early times ($t < 0.1\,$).  In the absence of neutrino
absorption, the distribution of electron fraction shifts to include additional
material with $Y_e < 0.1$ and exclude $Y_e > 0.4$. Nuclear recombination
deposits additional energy into the already unbound outflows at $t \sim 0.1\,$s,
resulting in ejecta with moderately higher velocities, but its absence only
decreases the total ejecta mass by 1\%. Inclusion of $r$-process heating by the
formation of heavier nuclei can further speed up the ejecta at late times
\citep{klion_2022}.  Proper characterization of the effect of neutrino
absorption and nuclear recombination on the mass ejection dynamics and
composition must be done with full 3D simulations, which unfortunately still
remain expensive computationally.

Our \emph{base} and \emph{base-tor} models have initial conditions consistent
with the post-merger system of the observed NS-NS merger GW170817. We find that
although these 3D models can eject lanthanide-free material ($Y_e > 0.25$) with
velocities inferred from kilonova modelling, $v \gtrsim 0.25\,c$, there is
insufficient mass in the outflows to match observations (e.g.,
\citealt{Kasen2017,Villar2017}) with the disk outflow alone.  The inclusion of a
finite-lived remnant in our \emph{base} model is a promising way to produce more
lanthanide-free ejecta at the required velocities (e.g.,
\citealt{FahlmanFernandez2018}). 

The main limitations of our work  are the approximations made for modelling
neutrino radiation transport on the necessary timescales.  Prior research into
the effectiveness of neutrino schemes have shown that the differences between
two-moment (M1) and Monte Carlo (MC) schemes can result in a $\sim 20$\%
uncertainty in neutrino luminosity, translating to a difference of 10\% in
outflow electron fraction \citep{Foucart2020MC}.  Comparison between M1 and
the leakage scheme of \citet{ILEAS2019} (which our scheme is based on, see
Appendix~\ref{AppC}) shows a further 10\% uncertainty in neutrino luminosities,
and a comparison of leakage+M0 to M1 schemes shows that leakage schemes tend to
decrease the average $Y_e$, but with a minimal effect on nucleosynthetic yields
\citep{Radice2021}.
The exclusion of relativistic effects implies that we
cannot accurately model jet formation, but the effects of these approximations
on mass ejection and  composition are likely minimal, since the relevant processes
operate far from the BH. The leading order special relativistic
corrections to the MHD equations are $\sim v/c$ for $v < c$, hence we estimate
uncertainties associated with our fastest ejecta to be at least of the same order
(e.g., 50\% for matter with $v_r/c =0.5$, etc.). The bulk of mass ejection has
$v_r/c < 0.1$, and thus uncertainties due to Newtonian physics should be on the
order of 10\%, comparable to those due to spatial resolution.

\section*{Acknowledgements}

We thank Sherwood Richers and Dmitri Pogosyan for helpful discussions.
This research was supported by the Natural Sciences and Engineering Research
Council of Canada (NSERC) through Discovery Grant RGPIN-2017-04286, and by the
Faculty of Science at the University of Alberta.  The software used in this work
was in part developed by the U.S. Department of Energy (DOE) 
NNSA-ASC OASCR Flash Center at the University
of Chicago.  
Data visualization was done in part using {\tt VisIt} \citep{VisIt}, which is supported
by DOE with funding from the Advanced Simulation and Computing Program
and the Scientific Discovery through Advanced Computing Program.
This research was enabled in part by support
provided by WestGrid (www.westgrid.ca), the Shared Hierarchical
Academic Research Computing Network (SHARCNET, www.sharcnet.ca),
Calcul Qu\'ebec (www.calculquebec.ca), and Compute Canada (www.computecanada.ca).
Computations were performed on the \emph{Niagara} supercomputer at the SciNet
HPC Consortium \citep{SciNet,Niagara}. SciNet is funded by the Canada
Foundation for Innovation, the Government of Ontario (Ontario Research Fund - Research Excellence),
and by the University of Toronto.

\section*{Data Availability}

The data underlying this article will be shared on reasonable request to the corresponding author.



\appendix

\section{MHD in Spherical Coordinates}
\label{AppA}

In this appendix we extend the work done in \citet{MSc}, where  the default
dimensionally-unsplit hydrodynamic solver in {\tt FLASH4} \citep{Lee2013} was
modified to work in non-uniform 3D spherical coordinates.  Here we add the
solution of the ideal MHD equations by including contributions at each face from
transverse and diagonal fluxes. To ensure the divergence-free evolution of
magnetic fields, \texttt{FLASH4.5} solves the induction equation via the
constrained transport (CT) method of \citet{Evans1988} on a uniform staggered
mesh. We also modify the default CT method to work in non-uniform 3D spherical
coordinates. 

\subsection{Governing Equations of Magnetohydrodynamics}
\label{sec:MHDOverview}

The ideal MHD equations are those of mass, momentum, and energy
conservation
(equations~\ref{eq:FLASHDensity}-\ref{eq:FLASHEnergy} {without source
terms})
together with the induction equation~(\ref{eq:Induction}).
Note that in {\tt FLASH} the induction equation is implemented as 
\begin{align}
    \label{eq:FLASHInduction}
    \der{\B}{t} + \div{(\B\otimes\v - \v\otimes\B)} = 0
\end{align}
which has the opposite sign as equation (\ref{eq:Induction}).
For the rest of this Appendix, we will use this form 
of the induction equation 
for consistency with the available literature on {\tt FLASH}.  

The conservation equations are written in terms of a vector of 
conserved variables $\U$, a tensor of associated fluxes $\mathcal F$, and a vector
of source terms $\mathbf S$,
 \begin{align}
    \label{eq:MHDEvolution}
    \pder{\U}{t} + \div{\mathF} = \mathbf{S}.
\end{align}
Writing out the associated fluxes in the radial, polar,
and azimuthal direction as $\mathbf{F}$, $\mathbf{G}$, and $\mathbf{H}$
respectively, we obtain
\begin{align}
    \label{eq:MHDcons}
    \pder{\U}{t} + \frac{1}{r^2}\pder{(r^2\F)}{r} + \frac{1}{r\sin\theta}\pder{(\sin\theta\G)}{\theta} + \frac{1}{r \sin\theta}\pder{\H}{\phi} = \mathbf S,
\end{align}
where \citep{Lee2013} 
\begin{align}
\label{eq:MHDFlux3} 
  \mathbf{U} =\begin{pmatrix}\rho          \\ \rho v_r          \\ \rho v_\theta
\\ \rho v_\phi \\B_r \\ B_\theta \\ B_\phi    \\ \rho E        \end{pmatrix},\qquad 
&\mathbf{F} =\begin{pmatrix}\rho v_r      \\ \rho v_r^2 + P - B_r^2   \\ \rho v_r
v_\theta - B_r B_\theta \\ \rho v_r v_\phi - B_r B_\phi\\ 0 \\ v_r B_\theta -
v_\theta B_r = -E_\phi \\ v_r B_\phi - v_\phi B_r = E_\theta \\ v_r(\rho E+ P) - B_r (\v
\cdot \B )\end{pmatrix}, \nonumber \\
\noalign{\smallskip}
& \mathbf{G} =\begin{pmatrix}\rho v_\theta \\ \rho v_\theta v_r - B_\theta B_r\\ \rho
v_\theta^2 + P - B_\theta^2 \\ \rho v_\theta v_\phi - B_\theta B_\phi  \\
v_\theta B_r - v_r B_\theta  = E_\phi \\ 0 \\ v_\theta B_\phi - v_\phi B_\theta =
-E_r \\ v_\theta(\rho E+ P) - B_\theta (\v \cdot \B )\end{pmatrix},\nonumber\\
\noalign{\smallskip}
&\mathbf{H} =\begin{pmatrix}\rho v_\phi   \\ \rho v_\phi v_r -  B_\phi B_r   \\
\rho v_\theta v_\phi - B_\theta B_\phi \\ \rho v_\phi^2 + P - B_\phi^2 \\ v_\phi
B_r - v_r B_\phi = -E_\theta \\ v_\phi B_\theta - v_\theta B_\phi  = E_r \\ 0  \\ v_\phi(\rho E+ P) - B_\phi (\v \cdot \B )\end{pmatrix}.
\end{align}

In conservative mesh-based codes like {\tt FLASH}, the conserved variables are
evolved by discretizing (\ref{eq:MHDcons}) in both time and space. This means
that we evolve each variable by taking a volume average over the cell, and then
advance the volume-averaged variable using the corresponding fluxes at the
faces. Using the indices $\{i,j,k\}$ to 
enumerate cells in the radial, polar, and azimuthal directions, respectively, 
using half integer indices to denote values at cell faces
($\pm \frac{1}{2}$), and using the index $n$ 
to denote time step, the conservative
system of equations can be written concisely as 
\begin{align}
    \label{eq:conupdate}
	\langle \U^{n+1}_{i,j,k}\rangle = \langle\U^n_{i,j,k}\rangle - \frac{\Delta t}{V} \bigg{[} &(A_{i-\frac{1}{2}} \F_{i-\frac{1}{2}} - A_{i+\frac{1}{2}} \F_{i+\frac{1}{2}}) \nonumber \\ 
- &(A_{j-\frac{1}{2}} \G_{j-\frac{1}{2}} - A_{j+\frac{1}{2}} \G_{j+\frac{1}{2}}) \nonumber \\ 
-&(A_{k-\frac{1}{2}} \H_{k-\frac{1}{2}} - A_{k+\frac{1}{2}} \H_{k+\frac{1}{2}}) \bigg{]} \nonumber\\
+  & \Delta t\langle\mathbf{S}\rangle,
\end{align}
where $A_{\{i,j,k\}\pm\frac{1}{2}}$ is the area of the cell face perpendicular to the denoted
direction, $V$ is volume of the cell, $\langle \mathbf{S} \rangle$ is the volume-averaged
source term, and $\Delta t = t^{n+1}-t^n$ is the timestep. We have used Gauss's Law, 
\begin{align}
  \label{eq:GaussLaw} 
  \iiint_V \div{\F} \mathrm{d}V = \oiint_{A} \F \cdot \mathrm{d} \mathbf{A},
\end{align}
to relate volume-averaged conserved
variables $\langle U \rangle$ to the face-centered fluxes. To make the
conservative variable update consistent with curvilinear coordinates, we 
compute the discretized cell face areas and volumes for arbitrarily spaced
spherical cells \citep{MSc}. 

\subsection{Geometric Source Terms in Spherical Coordinates}

The geometric source terms arise when taking covariant derivatives of second
rank tensors, often referred to as a tensor divergence. These terms take the physical form
of fictitious forces, and only arise in the equations
for vector quantities.
The scalar energy and density evolution equations therefore do not have source terms. For a tensor, $T$, in
spherical coordinates the divergence is written as (see \citealt{MSc}, or
\citealt{Mignone2005} for a separate derivation)
\begin{align}
\div{T} = \begin{pmatrix}
\nabla_r T \\ 
\nabla_\theta T \\ 
\nabla_\phi T
\end{pmatrix},
\end{align}
where we can define the divergences in each individual direction as
\begin{multline}  
\label{eq:rdiv}
\nabla_r T = \l\frac{1}{r^2}\pder{(r^2T^{rr})}{r}  + \frac{1}{r\sin\theta}\pder{(\sin\theta T^{\theta r})}{\theta} + \frac{1}{r\sin\theta}\pder{T^{\phi r}}{\phi} -\\ \frac{T^{\theta \theta} +  T^{\phi \phi}}{r} \r,  
\end{multline}
\begin{multline}
\nabla_\theta T = \l\frac{1}{r^2}\pder{(r^2T^{r\theta})}{r}  +\frac{1}{r\sin\theta} \pder{(\sin\theta T^{\theta \theta})}{\theta} + \frac{1}{r\sin\theta}\pder{T^{\phi \theta }}{\phi} +\\ \frac{ T^{\theta r}}{r} - \frac{T^{\phi \phi} }{r} \cot\theta \r, 
\end{multline}
\begin{multline}
\label{eq:phidiv}
\nabla_\phi T = \l\frac{1}{r^2}\pder{(r^2T^{r\phi})}{r}  + \frac{1}{r\sin\theta} \pder{(\sin\theta T^{\theta \phi})}{\theta} + \frac{1}{r\sin\theta}\pder{T^{\phi \phi }}{\phi} +\\ \frac{T^{\phi r}}{r} + \frac{T^{\phi \theta}}{r} \cot\theta \r.
\end{multline}
Explicitly applying the divergences (\ref{eq:rdiv})-(\ref{eq:phidiv}) to the
dyads in the momentum and induction equations (\ref{eq:FLASHMomentum} and
\ref{eq:Induction}) yields the source term vector in equation~(\ref{eq:MHDEvolution}) 
\begin{align}
\mathbf S = \begin{pmatrix}
      0 \\ 
     \dfrac{\rho (v_\theta^2 + v_\phi^2) - B_\theta^2 - B_\phi^2 }{r} \\[2ex]
    -\dfrac{\rho v_r v_\theta- B_r B_\theta}{r} + \dfrac{\cot{\theta}(\rho v_\phi^2-B_\phi^2)}{r} \\[2ex]
    -\dfrac{\rho v_r v_\phi - B_r B_\phi }{r} - \dfrac{\cot\theta(\rho v_\phi v_\theta -B_\phi B_\theta )}{r} \\
      0 \\  
    \dfrac{(B_\theta v_r - B_r v_\theta)}{r} \\[2ex]
    \dfrac{(B_\phi v_r - B_r v_\phi)}{r} + \dfrac{(B_\phi v_\theta - B_\theta v_\phi)\cot\theta}{r} \\
      0
  \end{pmatrix}.
\end{align}
The source terms have to be volume averaged according to
\begin{align}
\label{eq:s_vol_average}
    \langle \mathbf S \rangle = \frac{1}{V}\iiint_V \mathbf S(r, \theta) \d V.
\end{align}
for use in the discretized conservative update equation~(\ref{eq:conupdate}).
Since conservative mesh schemes store hydrodynamic variables as 
volume averages, we make the approximation of taking them out of the integral in equation~(\ref{eq:s_vol_average}). 
The remaining terms contain either a $r^{-1}$ or $\cot \theta$ dependence, which are
integrated analytically to get the volume-averaged source term vector
\begin{align}
\langle \mathbf S \rangle = \dfrac{3\Delta r^2}{2\Delta r^3}\begin{pmatrix}
      0 \\ 
     \rho (v_\theta^2 + v_\phi^2) - B_\theta^2 - B_\phi^2  \\[2ex]
    -\rho v_r v_\theta- B_r B_\theta + \dfrac{\Delta \sin\theta}{\Delta\cos\theta}(\rho v_\phi^2-B_\phi^2) \\[2ex]
    -\rho v_r v_\phi - B_r B_\phi  - \dfrac{\Delta\sin\theta}{\Delta\cos\theta}(\rho v_\phi v_\theta -B_\phi B_\theta ) \\[2ex]
      0 \\  
    (B_\theta v_r - B_r v_\theta) \\[2ex]
    (B_\phi v_r - B_r v_\phi) + \dfrac{\Delta\sin\theta}{\Delta\cos\theta}(B_\phi v_\theta - B_\theta v_\phi) \\[2ex]
      0
  \end{pmatrix},
\end{align}
where the coordinate differences are defined as
\begin{align}
  \Delta r^2 &= r^2_{i+\frac{1}{2}} - r^2_{i-\frac{1}{2}}, \\
  \Delta r^3 &= r^3_{i+\frac{1}{2}} - r^3_{i-\frac{1}{2}}, \\
  \Delta \cos\theta &= \cos\theta_{j+\frac{1}{2}} - \cos\theta_{j-\frac{1}{2}}, \\
  \Delta \sin\theta &= \sin\theta_{j+\frac{1}{2}} - \sin\theta_{j-\frac{1}{2}}.
\end{align}
These are the source terms used in the  update of 
conserved variables (equation~\ref{eq:conupdate}).

\subsection{Fluxes and Primitive Variables in Spherical Coordinates}
\label{sec:MHDprim}

Before the conservative update (equation~\ref{eq:conupdate}) can be performed, 
the fluxes at the face must be
known. Here we provide only the information needed to adjust the reconstruction
of fluxes for spherical coordinates.
The discretization and methods used in \texttt{FLASH} to extend
variables to the face and construct appropriate Riemann states can be found in
\citet{Lee2013}. By default, {\tt FLASH} employs a piecewise linear MUSCL-Hancock method
\citep{ColellaPLM1985} to reconstruct the so-called cell-centered primitive variables,
\begin{align}
    \V &= (\rho,v_r, v_\theta, v_\phi, B_r, B_\theta, B_\phi, \Pgas)^T,
\end{align}
instead of the conserved variables, to the faces.
The corresponding primitive system of equations is derived 
using the chain rule to expand the divergence in the conservative system.
The primitive system is utilized in part because it has a quasi-linear form,
which is written out in spherical coordinates as
\begin{align}
    \label{eq:primvarupdate}
    \pder{\V}{t} &+ \mathbf{M}_r\pder{\V}{r} + \mathbf{M}_\theta\frac{1}{r}\pder{\V}{\theta} + \mathbf{M}_\phi\frac{1}{r\sin\theta}\pder{\V}{\phi} = \mathbf{S}_p,
\end{align}
where the matrices $\M_r, \M_\theta, \M_\phi$ can be found in \citet{Lee2009} and \citet{Lee2013}. 
Explicitly expanding all the spatial and temporal derivatives of the
conseravative equations
(\ref{eq:FLASHDensity}, \ref{eq:FLASHMomentum}, \ref{eq:FLASHEnergy}, \ref{eq:FLASHInduction})
in spherical coordinates results in a source term vector for the primitive
system of variables given by
\begin{align}
\mathbf S_p = \begin{pmatrix} 
     -\rho\l \dfrac{2v_r + v_{\theta} \cot{\theta}}{r} \r  \\[2ex] 
     \dfrac{(v_\theta^2 + v_\phi^2) -(B_\theta^2 + B_\phi^2)/\rho }{r} \\[2ex]
    -\dfrac{v_r v_\theta- B_r B_\theta/\rho }{r} + \dfrac{\cot{\theta}(v_\phi^2-B_\phi^2/\rho)}{r} \\[2ex]
    -\dfrac{v_r v_\phi -  B_r B_\phi/\rho }{r} - \dfrac{\cot\theta(v_\phi v_\theta -B_\phi B_\theta/\rho )}{r} \\[2ex]
    -\dfrac{\cot\theta (B_\theta v_r - B_r v_\theta)}{r} \\[2ex]
    -\dfrac{(B_\theta v_r - B_r v_\theta)}{r} \\[2ex]
     \dfrac{(B_\phi v_r - B_r v_\phi)}{r}  \\[2ex]
     -\gamma_1 \Pgas\l \dfrac{2v_r + v_{\theta} \cot{\theta}}{r} \r  
  \end{pmatrix}.
\end{align}
These are the source terms implemented in the reconstruction of cell-centered
variables to the faces, along with the geometrically correct lengths, in equation~(\ref{eq:primvarupdate}). 

\subsection{Constrained Transport}
\label{sec:CT}

Magnetic fields require an update method different from 
conserved quantities, as the induction equation does not explicitly require
that the solenoidal constraint,
\begin{align}
  \div{\B} = 0,
\end{align}
{is}
satisfied. 
To satisfy this constraint,
{fields are taken to be averaged over
cell faces}, 
and the induction equation is written as
\begin{align}
    \oiint \pder{\B}{t} \cdot \d \mathbf{A} &= \oiint -\nabla \times \mathbf{E} \cdot \d\mathbf{A}.
\end{align}
Applying Stokes' theorem, using the definition of magnetic fields as 
area-averaged, and discretizing the resulting line integral as a sum, we arrive at the
Newtonian form of CT \citep{Evans1988}
\begin{align}
    \label{eq:CT}
    \B \cdot \mathbf{A} &= -\frac{\Delta t}{2} \sum_\mathrm{edges} \mathbf{E} \cdot \Delta \boldsymbol \ell.
\end{align}
To perform the summation, we need the
electric fields around each edge of the cell. These are found by Taylor
expanding the face centered electric fields, 
which are obtained from the fluxes of conserved magnetic variables
(equation~\ref{eq:MHDFlux3}) in the reconstruction step, and using Ohm's law in ideal MHD
\begin{align}
\label{eq:ohm_ideal_mhd}
  \E = - \v \times \B.
\end{align}
\begin{figure}
\includegraphics*[width=0.48\columnwidth]{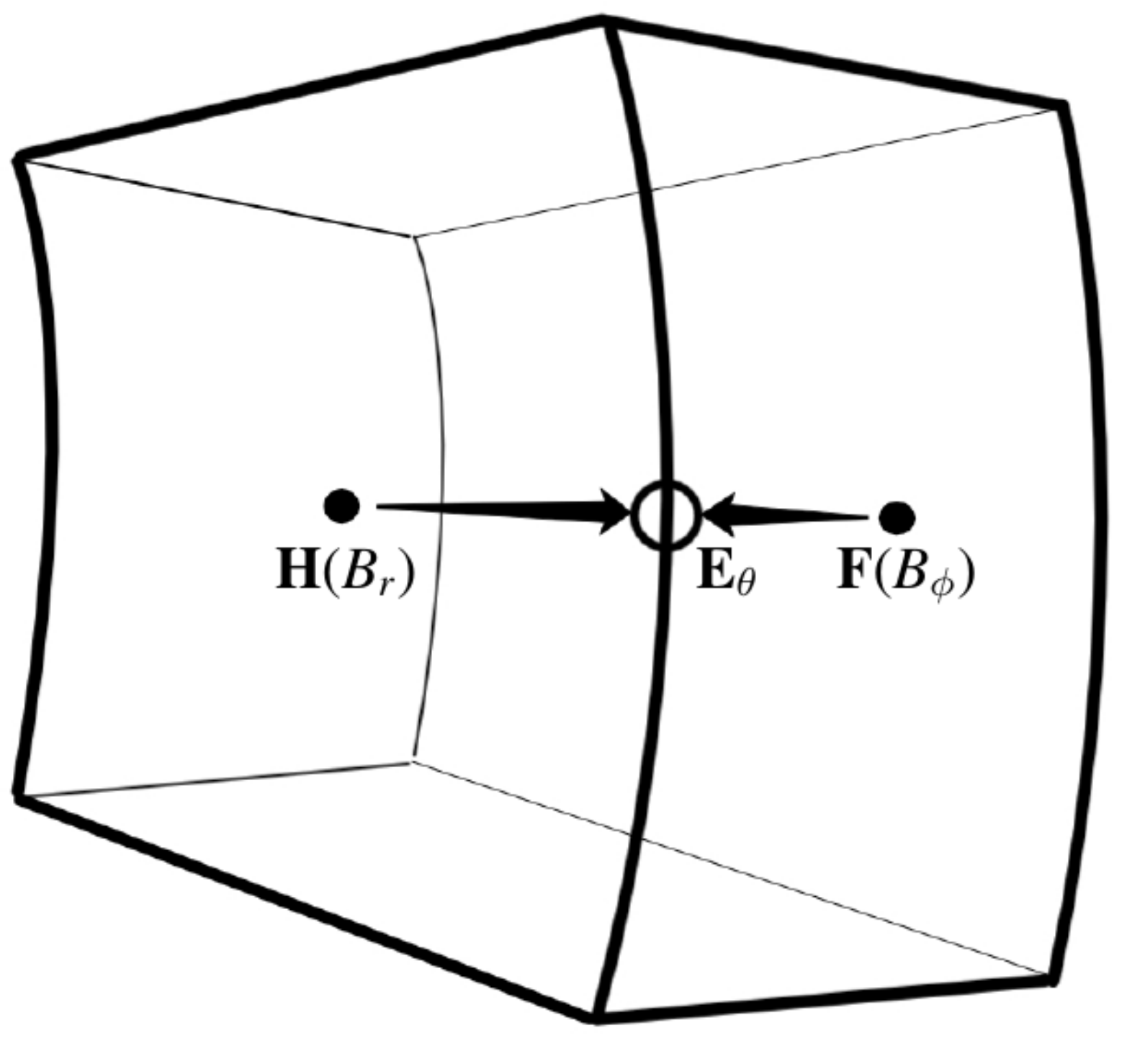}
\includegraphics*[width=0.52\columnwidth]{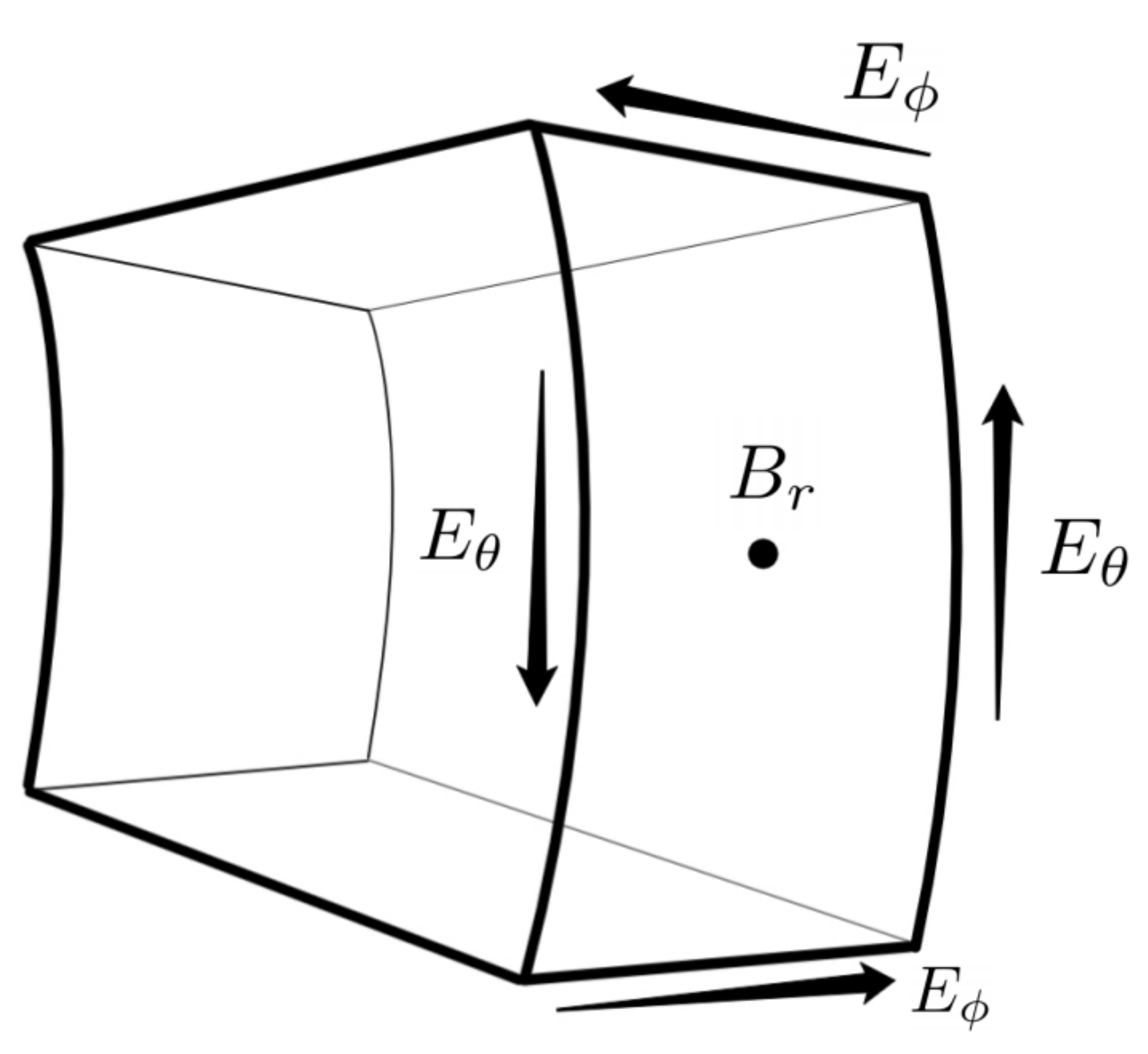}
\caption{\emph{Left:} Schematic representation of the  
method to obtain the electric field at cell edges for the CT method
(equation~\ref{eq:CT}).  The electric fields at cell faces are obtained using
Ohm's law in ideal MHD (equation~\ref{eq:ohm_ideal_mhd}) from the magnetic
fluxes at the face (solid black circles). The electric fields are then
Taylor-expanded to the cell edge using finite derivatives
(equations~[\ref{eq:FDReal}]-[\ref{eq:FD2Real}]).  The expanded values are then
averaged using upwinding  for numerical stability (open circle, see text).  Two
additional symmetric expansions to the same edge have been left out for visual
clarity.  \emph{Right:} Schematic description of the reconstruction of the
face-centered radial magnetic field by adding up the total contribution from
electric fields along each of the cell edges (equation~\ref{eq:CT}).}
\label{fig:EfieldExpansion}
\end{figure} 
Each flux contains 2 electric fields, one in each transverse direction. At each
edge there will be 4 independent fluxes that can be used to find the electric
field at that point (Figure~\ref{fig:EfieldExpansion} {shows two of the
fluxes})
Conventionally, all 4 contributions
are included, which can be numerically unstable 
(e.g., \citealt{Toth2000,Mignone2020}).  Fortunately, {\tt FLASH} includes an
option to only use the upwind fluxes (flux vectors which have a positive mass
flux) to construct the electric fields (e.g., \citealt{White2016}). We find that
this upwinded method is necessary to preserve numerical stability when  using
{\tt FLASH} to evolve a magnetized torus in 3D spherical coordinates, despite
not being a default option. This is also found by \citet{Kuroda2020}, who note
that in the regions of their core-collapse supernova setup where matter is
supersonically advecting (analogous to the torus), the upwind method is
necessary for stability. While no additional work is needed to implement the
upwinding in spherical coordinates, the Taylor expansion for expanding the face
fluxes to cell edges must be updated to account for non-uniform grid spacing.
The Taylor expansion uses second-order finite differences to determine the
approximate value at face edges. For example, the expansion in an arbitrary
$i^\mathrm{th}$ direction is 
\begin{align}
\label{eq:taylor_cell_edge}
   E_{i\pm\frac{1}{2},j,k} = E_{i,j,k} \pm \Delta \ell_{i\pm\frac{1}{2}} \pder{E}{\ell} +
(\Delta \ell_{i\pm\frac{1}{2}})^2\frac{1}{2}\pder{^2E}{\ell^2},
\end{align} 
where $\Delta\ell_{i\pm1/2}$ corresponds to the distance
from the center of the cell to the $i^\mathrm{th}$ face.
Writing the spatial derivatives as cell-centered finite differences, where
$\Delta \ell_\pm$ is the distance 
from the center of the cell $i$ to the cell centers above and below ($i\pm1$),
yields 
\begin{align}
  \label{eq:FDReal}
  \pder{E}{\ell}  &= \frac{E(\ell + \Delta \ell_+) - E(\ell - \Delta \ell_-) }{\Delta \ell_+ + \Delta \ell_-} \\
  \label{eq:FD2Real}
  \pder{^2E}{\ell^2} &= \dfrac{\dfrac{E(\ell + \Delta \ell_+) - E(\ell) }{\Delta
\ell_+} - \dfrac{E(\ell) - E(\ell - \Delta \ell_-) }{\Delta \ell_-}}{\Delta \ell_- + \Delta \ell_+}.
\end{align} 
For uniform spacing, $\Delta \ell_+ = \Delta \ell_- = \Delta \ell$, and the first and second order derivative finite differences simplify to
\begin{align}
  \label{eq:FDFlash}
  \pder{E}{\ell}  &= \frac{E(\ell + \Delta \ell) - E(\ell - \Delta \ell) }{2\Delta \ell} \\
  \label{eq:FD2Flash}
  \pder{^2E}{\ell^2} &= \frac{E(\ell + \Delta \ell) - 2E(\ell) + E(\ell - \Delta \ell) }{2(\Delta \ell)^2}.
\end{align} 
Since the public version of {\tt FLASH}  
only supports a uniform grid,
the equations in the code (\ref{eq:FDFlash}-\ref{eq:FD2Flash}) 
must be modified so that the more general case
(\ref{eq:FDReal}-\ref{eq:FD2Real}) is used for non-uniform grid spacing. 
Once the electric field construction is complete, the magnetic fields are
updated to half timestep with equation (\ref{eq:CT}) by adding up
electric fields around the cell faces (Figure~\ref{fig:EfieldExpansion}).

\section{Tests of MHD in \texttt{FLASH}} 
\label{AppB}
In order to demonstrate the accuracy of our extension of the unsplit MHD module to non-uniform 3D
spherical
coordinates, we present the results of 
two strong MHD tests: a magnetized blast wave and
a magnetized accretion torus. Although neither of these tests has an analytic
solution, we compare our results numerically to those 
obtained with the well-tested
cartesian and cylindrical MHD implementations in {\tt FLASH}. 
In all of our spherical setups, we use a
grid that is evenly spaced in $\cos\theta$ in the polar direction to increase
our timestep in the polar regions. All tests use an HLLD Riemann solver and the
piecewise linear reconstruction method.

\subsection{Magnetized Blast Results} 

The magnetized blast wave is an extension of the classic hydrodynamic Sedov
blast wave to MHD (e.g., \citealt{Komissarov1999}). An initially overpressurized
region $P\sim500\,\mathrm{erg\,cm}^{-3}$ encompassing a few cells at the inner boundary
($r < 0.125\,\mathrm{cm}$) is allowed to expand into a uniform lower pressure
$P=1\,\mathrm{erg\,cm}^{-3}$ medium.  The domain is initialized with uniform density
$\rho=1\,\mathrm{g}\,\mathrm{cm}^{-3}$, as well as a uniform magnetic field in
the cartesian $x-$direction. {Due to magnetic tension}, 
expansion of the blast wave is inhibited in the direction parallel to the
$x$-axis, resulting in an asymmetric explosion. We use a dynamically important
initial magnetic field strength, $B_0 = 10 \,\mathrm{G}$, to test the code in
highly magnetized regions 
We perform the blast in both 3D cartesian and
3D spherical coordinates.  

\begin{figure}
\centering
\includegraphics[width=\columnwidth]{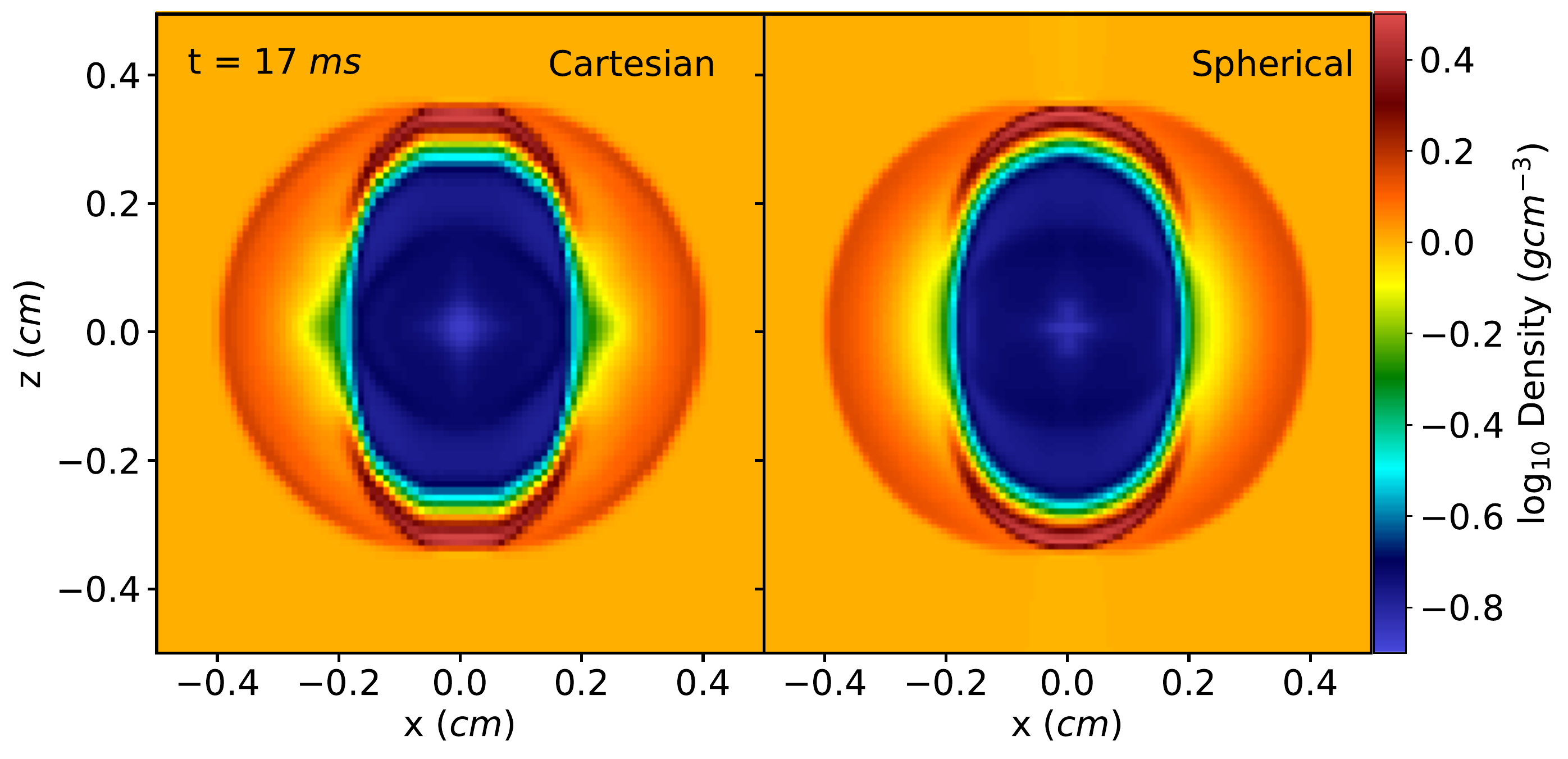}
\caption{Density slice at 17\,ms in $xz$-plane of the 3D-MHD blast wave in
Cartesian (left) and spherical (right) coordinates. A slice along the
$z$-direction is shown in Figure \ref{fig:MHDBlastComparison}.}
\label{fig:MHDBlastSnap}
\end{figure} 

Both simulations produce the same overall expansion 
of the blast-wave, as shown in Figure
\ref{fig:MHDBlastSnap}. The only large quantitative difference 
occurs at the origin, where the reflecting inner radial boundary condition causes small differences in
magnetic field evolution. In the cartesian explosion, the field is continuous
across $x=0$ and not reflecting. In {\tt Athena++}, \citet{Skinner2010} perform an 
off-center explosion to remove this difference, but we choose not to do so here 
due to the large resolution requirements. Also, an off-center explosion  
does not take advantage of the symmetry of the coordinate system. 
The cartesian test also shows visible edges,
which can be attributed to the shock being steeper in the spherical geometry.     

\begin{figure}
\centering
\includegraphics[width=1.1\columnwidth]{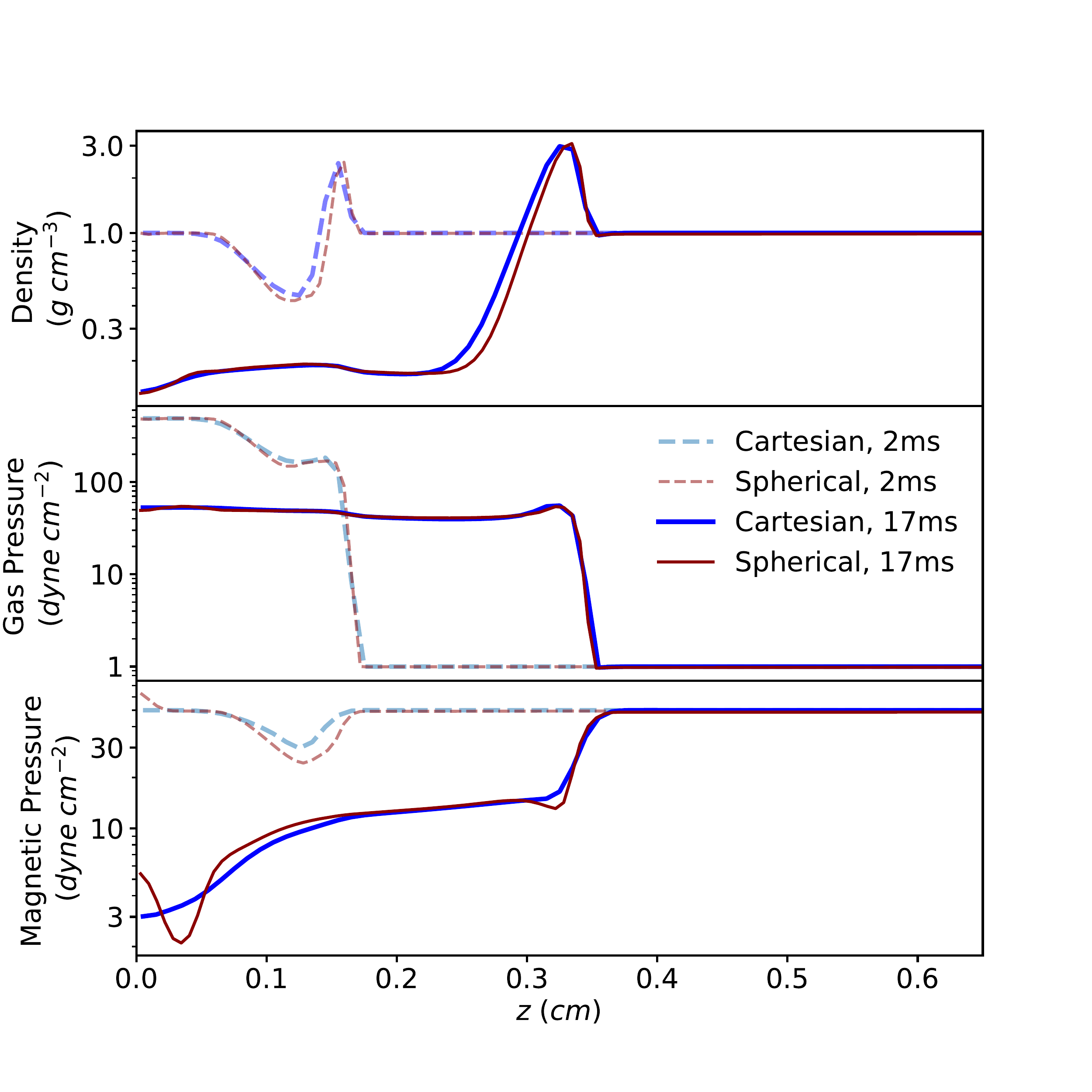}
\caption{Density slices along the $z-$direction ($x=0$, $y=0$) in the
3D MHD blast wave test, with geometries and times as labeled (see also Figure~\ref{fig:MHDBlastSnap}).
Notable differences appear in the magnetic pressure throughout the entire evolution due to the difference in boundary conditions between the two simulations. Other differences can be attributed to
the geometry and resolution.}
\label{fig:MHDBlastComparison}
\end{figure} 

We numerically compare a slice along the Cartesian $z-$direction in Figure
\ref{fig:MHDBlastComparison}. The two differences described previously are also noticeable in
these plots, especially in the magnetic field at the inner boundary. However,
both codes track the shock position identically, and the only 
noticeable differences appear at sharp gradients.  

\begin{figure}
\centering
\includegraphics[width=\columnwidth]{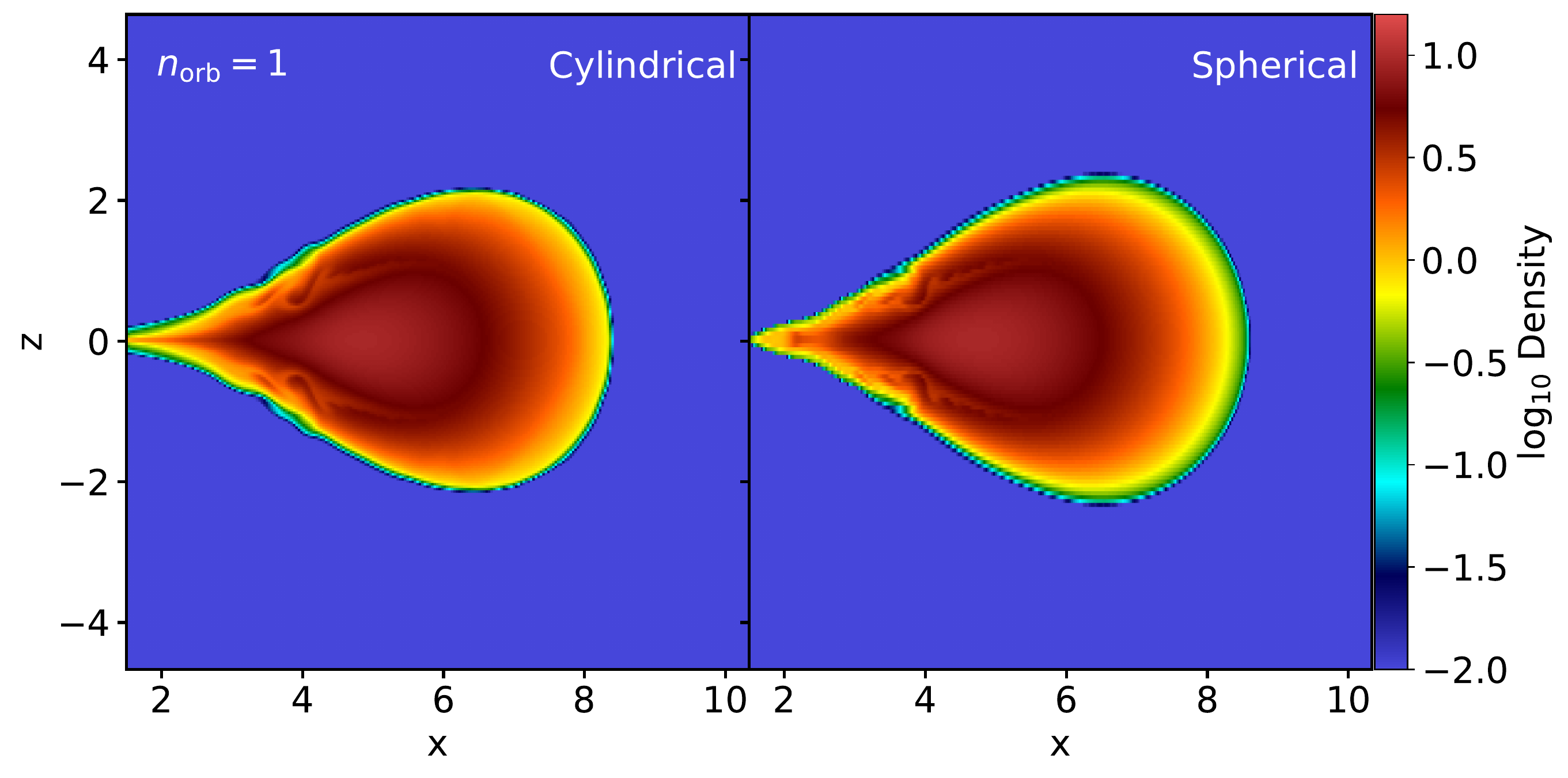} 
\caption{Density snapshot of
the 2.5D axisymmetric torus test after one orbit at the initial density peak
radius. The density field in cylindrical (left) and spherical (right)
coordinates is normalized to code units, with a maximum value of 10. Noticeable
differences appear in regions near the origin, where the flow is slightly more
resolved in spherical coordinates. }
\label{fig:2DTorusSnap}
\end{figure} 

\begin{figure}
\centering
\includegraphics[width=1.1\columnwidth]{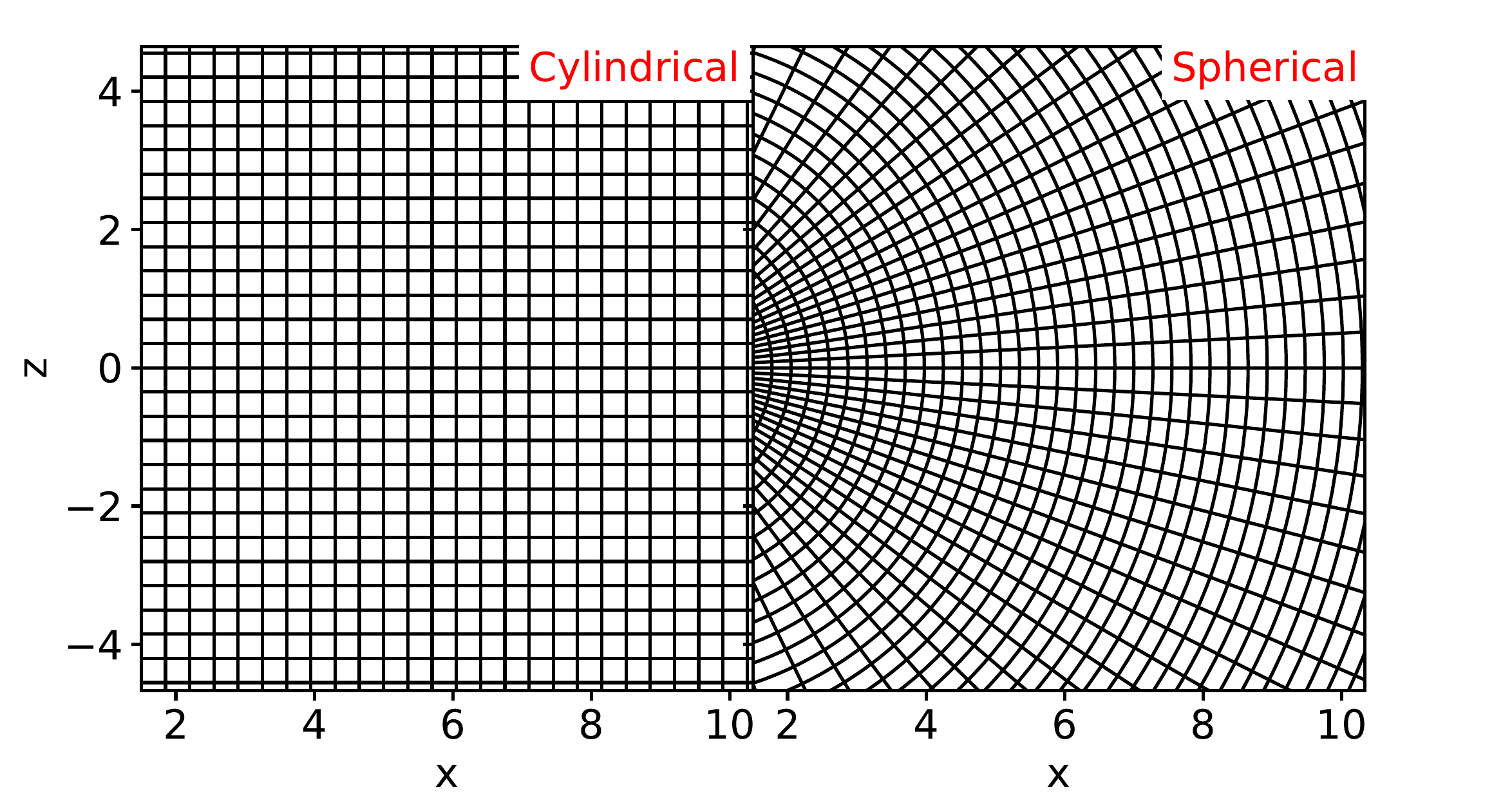}
\caption{Comparison of the grid setup in cylindrical and spherical coordinates 
for the 2.5D torus test runs. Every 10$^\mathrm{th}$ grid line is plotted for visual clarity. The
spherical grid is more resolved towards the inner radial part of the midplane,
but at comparable resolutions in the rest of the torus. The outer polar regions
are less resolved to reduce constraints on the simulation time step.}   
\label{fig:2DTorusGrid}
\end{figure} 

\begin{figure*}
\centering
\includegraphics[width=0.8\textwidth]{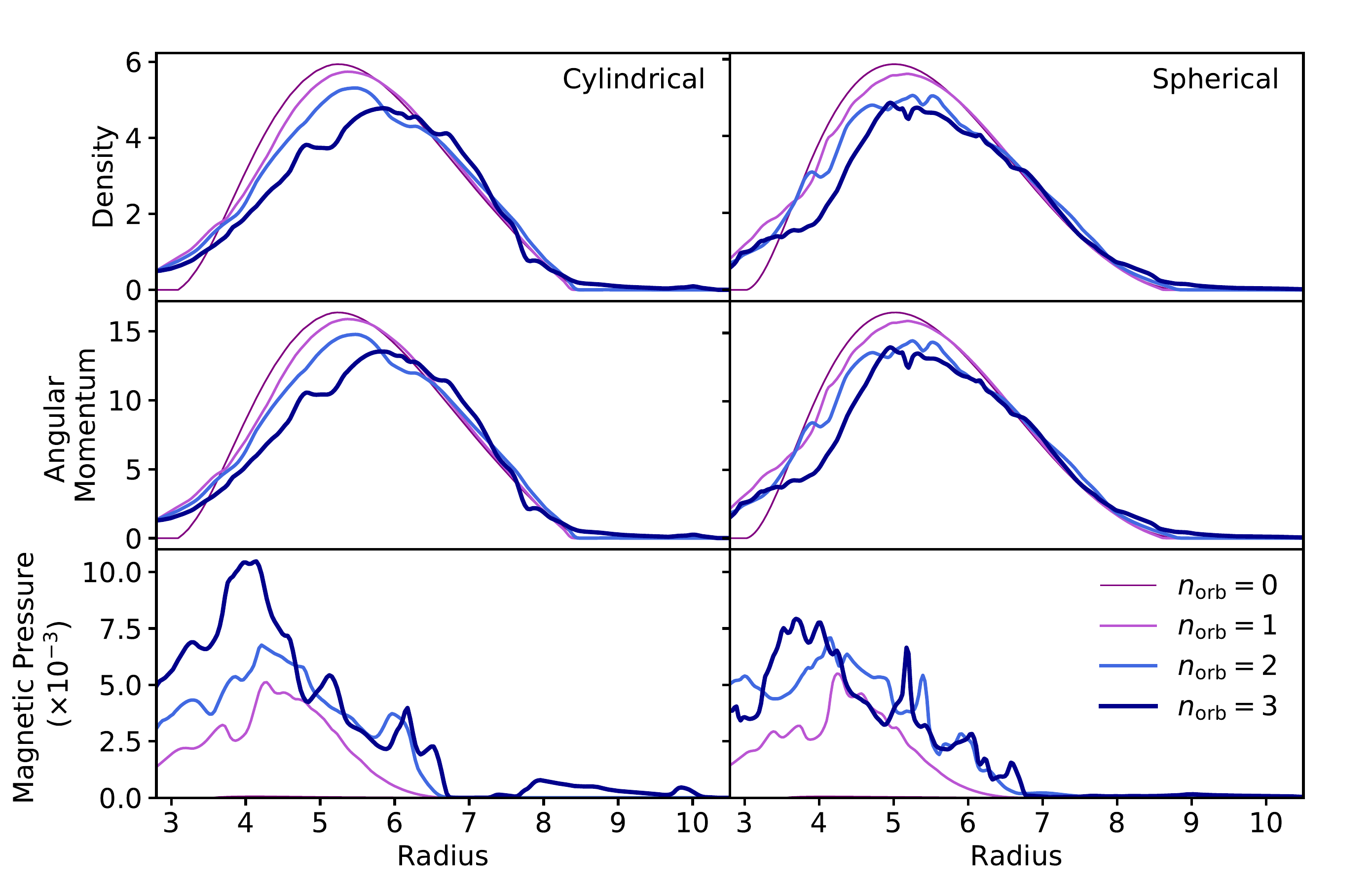}
\caption{Results from the 2.5D axisymmetric torus tests, evolved for 3 orbits at
the initial radius of maximum density $r_{\rm circ}$.  The left column shows the
height-averaged density, angular momentum, and magnetic pressure (all in code
units) in the cylindrical simulation, while the right column shows the same
quantities for the spherical case but angle-averaged instead.
Lines of increasing thickness show the torus evolution from
its initial to final state at each orbit ($n_{\rm orb}$). Numerical differences
between the angle- and height-averaging appear even at the 0th orbit, but the
general trends are not affected.} \label{fig:2DTorusMHDComparison}
\end{figure*} 

\subsection{Magnetized Torus Results}

We perform two torus tests. The first is carried out in 2.5D in both cylindrical
and spherical coordinates, and employs a standard and normal evolution (SANE)
initial magnetic field configuration. Results can be compared to multiple
previous implementations \citep{Hawley2000,Mignone2007,Tz2012}. The second test
is the extension to 3D spherical coordinates of the same SANE torus.

We set up the tori as in \citet{MSc}, using the gravitational potential of
\citep{PW1980} and setting the gravitating mass, $M=1$. We normalize units such
that $G = c = 1$. We choose the same parameters as
model GT1 from \citet{Hawley2000}, which creates a thin, constant angular
momentum torus with a maximum density of 10 and an orbital timescale of $\sim
50$ at the initial circularization radius $r_\mathrm{circ} = 4.7$, and a minimum
radius of $r=3$. The initial vector potential follows the
density distribution,
\begin{align}
  A_\phi = \max(\rho-\frac{1}{2}\rho_\mathrm{max},0),
\end{align}
which creates poloidal field loops threaded well within the torus. Initially, the
field is normalized to $\la \beta \ra = 100$, so the
field is dynamically unimportant and does not disturb the equilibrium condition 
The tori are then evolved for a total of $\sim3$ orbits at $r_{\rm circ}$. 

\begin{figure*}
\centering
\includegraphics[width=0.8\textwidth]{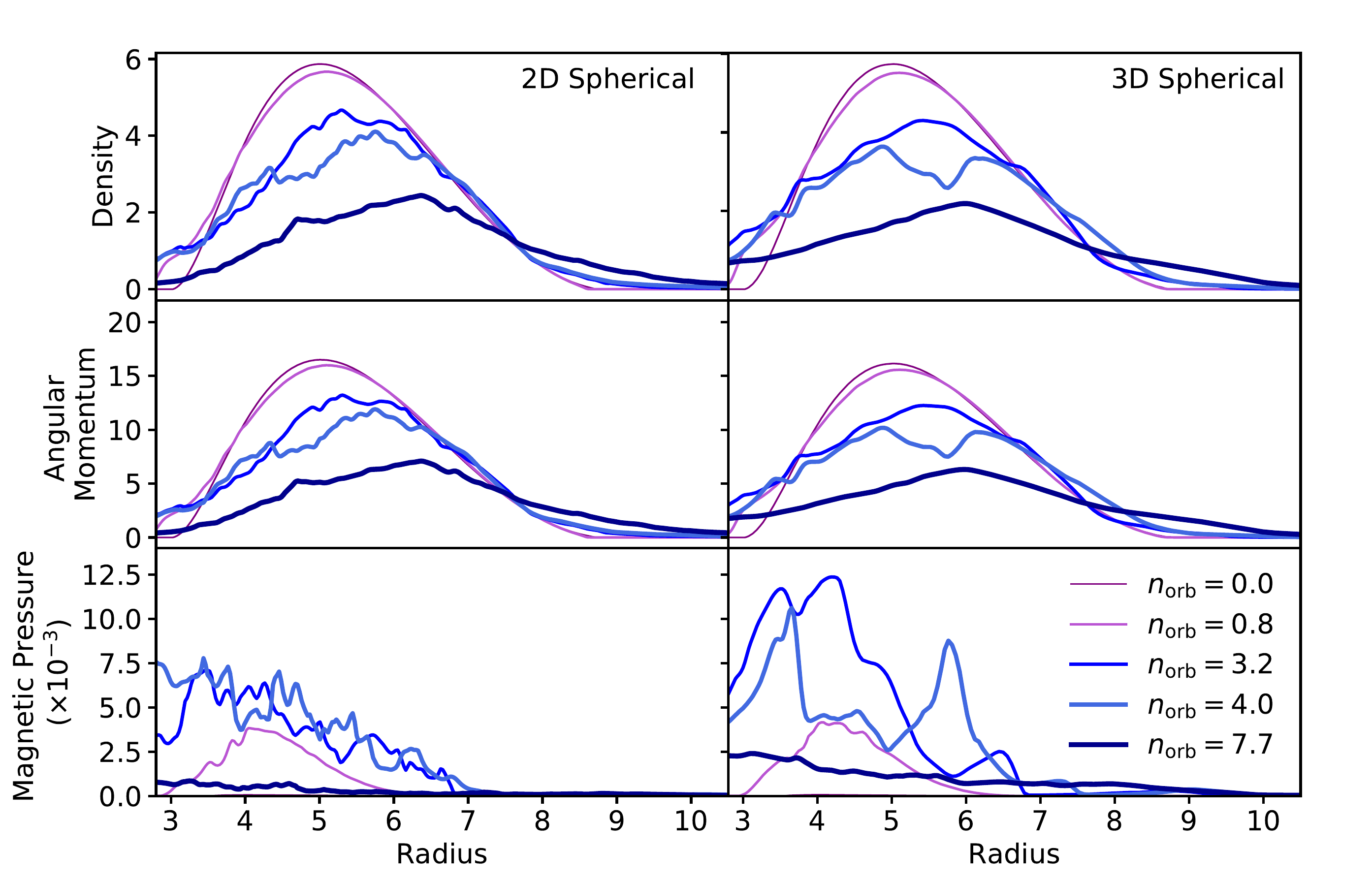}
\caption{Comparison between the 2.5D axisymmetric 
and 3D torus
test in spherical coordinates, showing the same angle-averaged quantities
as in Figure~\ref{fig:2DTorusMHDComparison} at various times. Both 
runs follow the expected evolution: the 3D torus sustains a
more powerful MRI for a longer time due to the additional azimuthal
turbulence, with the evolution being otherwise 
identical during the first $\sim$2 orbits.
Differences in spatial resolution manifest as smoother profiles in the 3D runs, which
has coarser cell sizes.}   
\label{fig:3DTorusMHDComparison}
\end{figure*} 

We compare the
evolution of hydrodynamic variables in the 2.5D runs quantitatively and qualitatively in
Figures~(\ref{fig:2DTorusSnap}-\ref{fig:2DTorusMHDComparison}). One notable
difference between the two axisymmetric runs is the resolution and the inner
boundary. In cylindrical coordinates, the entire inner $z-$boundary is set to be
absorbing, while in spherical coordinates the inner radial boundary is set to be
absorbing, and the polar boundaries are reflecting. This affects the dynamics
near the inner edges of the torus, which is noticeable in the
magnetic fields. The resolution in the two runs is comparable, but not exactly
the same as necessitated by the difference in grid structures. The spherical run
resolves the torus better, with $\Delta r = 0.028$ and the highest angular resolution in
the midplane, $r_\mathrm{circ}\min(\Delta\theta) = 0.024$, 
resulting in nearly square cells at that location, compared to the 
constant cylindrical resolution of $\Delta r_\mathrm{cyl} = 0.035$ and $\Delta z = 0.035$. 
The two grids are shown in Figure~\ref{fig:2DTorusGrid} for reference.

The axisymmetric runs follow the same qualitative evolution in cylindrical and
spherical coordinates: the azimuthal magnetic field grows due to winding, and
the magnetic pressure causes the torus to expand and accrete. The radial angular
momentum profile flattens as matter approaches the ISCO, and the magnetic field
grows largest in the central regions and then accretes, frozen in with the mass
flow. The first large divergences between the two simulations begin to appear
in the magnetic field  after $\sim 1$ orbit at $r_{\rm circ}$, when the
accretion stream reaches the inner boundary. Feedback from the reflecting
boundary changes the expansion of the torus into the ambient between the two
cases. Furthermore, additional turbulent structures manifest in the spherical
torus, noticeable as less smooth profiles in
Figure~\ref{fig:2DTorusMHDComparison}. We attribute this in part to the
differences in spatial resolution, and also note that \citet{Mignone2007} see
similar effects in their {\tt PLUTO} code tests with spherical and cylindrical
coordinates and the exact same setup (see their Figure 8). 

The 3D SANE torus follows the same evolution as the
axisymmetric spherical case for the first few orbits,
as shown in Figure~\ref{fig:3DTorusMHDComparison}. Notable differences begin to
appear after $\sim$3 orbits at $r_{\rm circ}$, as the MRI begins to die down in
the axisymmetric run. In 3D, the MRI creates stronger magnetic
fields and is sustained for longer through the additional turbulence in the
azimuthal direction \citep{Hawley2000}.  The profiles in the 3D run 
appear smoother, as we run it with a lower
resolution than in the 2D case due to computational 
limitations.  To make up for this difference in resolution,
we use a logarithmic grid in radius in the 3D model, so
that the region containing the torus is still resolved well. This corresponds to
resolutions in the radial,
polar, and azimuthal directions being $\Delta
r_\mathrm{circ} \sim 0.040$, $r_\mathrm{circ}\min(\Delta\theta) = 0.10$, and
$\Delta \phi = 0.03$.

\section{Neutrino Leakage Scheme} 
\label{AppC}
Neutrinos change the composition of disk outflows through 
charged current weak interactions
that alter the ratio of protons to neutrons. These transformations proceed via
emission or absorption of electron neutrinos and antineutrinos. 
A common approach to modeling emission of neutrinos is the 
so-called ``leakage scheme" 
\citep{Ruffert1996}, which interpolates between the diffusive and transparent
regimes {of radiative transport}.
Leakage schemes have been shown to capture the dominant
effects of neutrinos in post-merger tori around compact
objects, especially for BH disks, for which they are subdominant energy
sources \citep{Foucart2019, FM13, Siegel2018,Fernandez2019}. However, significant
differences appear when compared quantitatively 
to more advanced Monte-Carlo or two-moment (M1)
schemes \citep{Richers2015,Foucart2015,Perego2016,ILEAS2019,Radice2021}. For this reason it
is necessary to make impovements to the previous leakage-scheme implemented in
\texttt{FLASH} \citep{FM13,MF14}, while retaining computational efficiency. 

\subsection{Leakage Overview} 

The key components of a leakage scheme are the two source terms  that describe
the effective neutrino energy and number loss rate per unit volume for each
neutrino species,
\begin{align}
  \label{eq:Qruff}
  {Q}^{\rm eff}_\nui = Q_\nui \chi_{\nui,E}, \\ 
  \label{eq:Rruff}
  {R}^{\rm eff}_\nui = R_\nui \chi_{\nui,N}, 
\end{align} 
where $i$ represents a species ($\nu_e$ or $\bar{\nu}_e$ in our scheme), the
subscripts $E$ and $N$ refer to energy and number, respectively.  $Q$ and $R$
are the energy and number production rates per unit volume, respectively, which
are obtained from  analytic expressions \citep{Ruffert1996}. {The
scaling factors} $\chi_{\nui,E}$ and $\chi_{\nui,N}$ 
interpolate between the free-streaming and optically thick (diffusive)
regimes for neutrinos in both energy and number,
\begin{align}
\label{eq:leakage_factor}
  \chi_{\nui,\{E,N\}} = \l 1 + \frac{t^\mathrm{diff}_{\nui,\{E,N\}}}{t^\mathrm{loss}_{\nui,\{E,N\}}} \r^{-1},
\end{align}
where $t^\mathrm{diff}_\nui$ and $t^\mathrm{loss}_\nui$ are the diffusion and
loss timescales for each species, respectively.  The source terms for
equations~(\ref{eq:FLASHEnergy})-(\ref{eq:FLASHLepton}) are then obtained as
follows
\begin{eqnarray}
Q_{\rm net} & = & Q^{\rm eff}_{\nu_e} + Q^{\rm eff}_{\bar{\nu}_e} + Q_{\rm abs}\\
\Gamma_{\rm net} & = & \frac{m_n}{\rho}\left(R^{\rm eff}_{\nu_e} + R^{\rm eff}_{\bar{\nu}_e} \right) + \Gamma_{\rm abs},
\end{eqnarray}
where $Q_{\rm abs}$ and $\Gamma_{\rm abs}$ are the contributions from neutrino
absorption (treated separately) and $m_n$ is the neutron mass.  The diffusion
and loss timescales in equation (\ref{eq:leakage_factor}) are central to the
accuracy of the scheme, and thus we will discuss them in more detail.

\subsubsection{Loss timescale} 
Once the direct energy and number production rates in
(\ref{eq:Qruff}-\ref{eq:Rruff}) are found, the loss times are obtained as
\begin{align}
  t^\mathrm{loss}_{\nui,E} = \frac{Q_\nui}{E_\nui} \\ 
  t^\mathrm{loss}_{\nui,N} = \frac{R_\nui}{N_\nui},
\end{align}  
where $E_\nui$ and $N_\nui$ are the neutrino energy density and number density,
respectively. These quantities are obtained using analytic fits to Fermi
integrals from \citet{Takahashi1978}. 

\begin{figure}  
\centering
\includegraphics[width=\columnwidth]{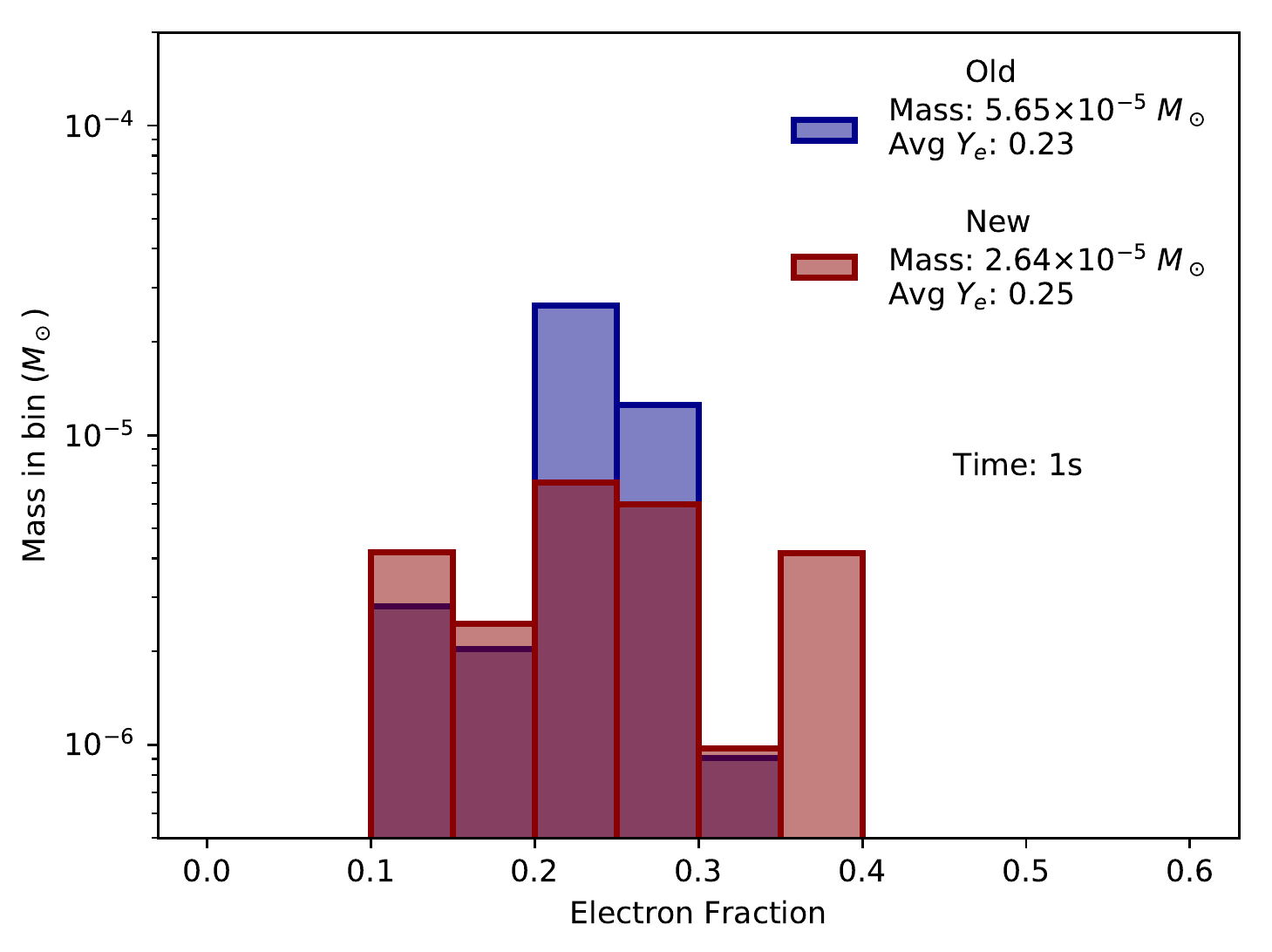}
\caption{Outflow mass histogram as a function of electron fraction after 1\,s of
evolution from two axisymmetric hydrodynamic BH disk setups that employ the old
and new implementation of diffusion time in the neutrino leakage scheme
(equations~\ref{eq:tdiff_simple}-\ref{eq:scale_height} and \ref{eq:tdiffEILEAS}
from \citealt{ILEAS2019}, respectively).  The average electron fraction of the
outflow increases by 5\%, and the outflow quantity decreases by about $50\%$ in
the new scheme, as the diffusion time correspondingly decreases by an order of
magnitude. This difference in outflow properties comes from the high velocity
neutrino-driven, early time outflows, which retain less energy deposited by
neutrinos, since cooling is more efficient.}
\label{fig:FLASH3TorusComparison}
\end{figure}

\subsubsection{Diffusion timescale}
The diffusion timescale is approximately given by
\begin{align}
\label{eq:tdiff_simple}
  t^\mathrm{diff}_{\nui,\{E,N\}} \sim \frac{3\kappa_{\nui,\{E,N\}} d^2}{c}.
\end{align}
where $\kappa_{\nui,\{E,N\}}$ is the energy or number opacity for species $i$,
and $d$ is a characteristic diffusion distance. A more accurate expression
involves calculation of the optical depth in various directions, which many
leakage schemes incorporate (e.g., \citealt{rosswog_2003}), but which is a
global calculation that is computationally expensive.  Our previous leakage
implementation \citep{FM13,MF14} approximates $d$ as the pressure scale height
assuming hydrostatic equilibrium in the cylindrical z-direction, which is the
preferential direction for neutrinos to escape the torus,
\begin{align}
\label{eq:scale_height}
  d \approx \frac{P}{\big{(}\pder{P}{r}\big{)}} = \frac{P}{\rho |\cos\theta g|}.
\end{align}
This is a local calculation which yields a neutrino optical depth
correct to within a factor of $\sim 2$.

Recently, \cite{ILEAS2019} have developed a novel method for determining the
diffusion timescale which is local (computationally efficient) and accurate.  In
this method, the diffusion timescale is determined from the diffusion equation
using a flux limiter
\begin{align}
  \label{eq:Etransport}
  \pder{\{E,N\}_\nui}{t} = -\div{\mathbf{F}_{\nui,\{E,N\}}}
\end{align}
where $\mathbf{F}_{\nui,\{E,N\}}$ is the neutrino energy of number flux,
 respectively. In flux-limited diffusion
\citep[FLD, see e.g.,][]{Wilson1975,Levermore1981,Kolb2013},
 the flux is given by 
\begin{align}
  \label{eq:FicksE}
  \mathbf{F}_{\nui,\{E,N\}} = -\frac{c}{3\kappa_\nui} \Lambda_{\nui,\{E,N\}}\nabla \{E,N\}_\nui, 
\end{align}
where $\Lambda_{\nui,\{E,N\}}$ is a flux limiter that interpolates between pure
diffusion ($\Lambda_{\nui,\{E,N\}} = 1$) and free-streaming (${F}_{\nui,\{E,N\}}
= c\{E_\nui,N_\nui\}$).  The diffusion times can be determined analogously to
the loss timescales:
\begin{align} 
  \label{eq:tdiffFicksE}
  t^\mathrm{diff}_{\nui,\{E,N\}} =\frac{\{E,N\}_\nui}{\l\dpder{\{E,N\}_\nui}{t}\r}.
\end{align}  
Expanding out the time derivative using the diffusion equations
(\ref{eq:FicksE}) and energy/number transport (\ref{eq:Etransport}) yields
\begin{align}
  \label{eq:tdiffEILEAS}
  t^\mathrm{diff}_{\nui,\{E,N\}} = \frac{\{E,N\}_\nui}{\div{(\frac{-c}{3\kappa_\nui}\Lambda_{\nui,\{E,N\}} \nabla \{E,N\}_\nui})}. 
\end{align}
We follow \citet{ILEAS2019} in using the flux limiter of
\citet{Wilson1975} for each species,
\begin{align}
  \label{eq:FLDLimE}
  \Lambda_{\nui,\{E,N\}} = \l 1 + \frac{1}{3\kappa_\nui}\frac{|\nabla \{E,N\}_\nui|}{\{E,N\}_\nui} \r^{-1}, 
\end{align}
In contrast to \citet{ILEAS2019} who integrate quantities over the neutrino
distribution, we use energy-averaged (over a Fermi-Dirac distribution) opacities, energy densities, and number
densities in equation~(\ref{eq:tdiffEILEAS}), computing only the spatial
gradient. 
\begin{figure*}
\centering
\includegraphics[width=\textwidth]{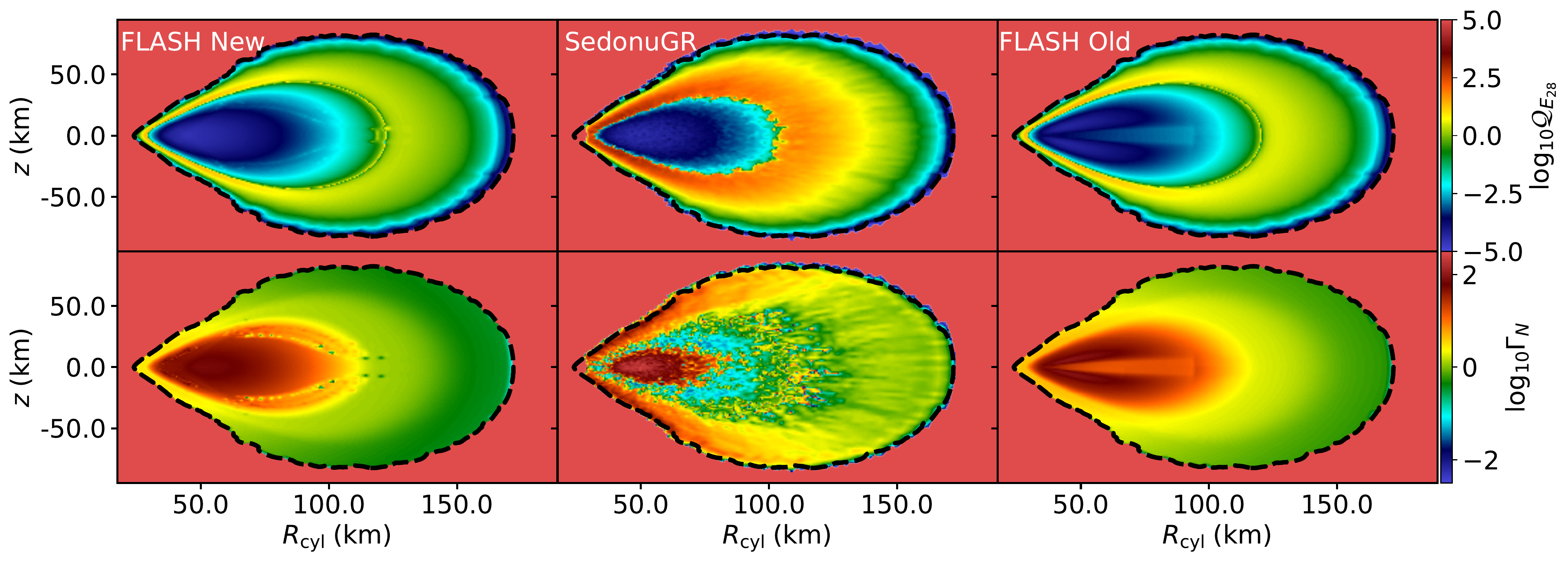}
\caption{Comparison of neutrino source terms 
{(\emph{Top}: energy rate per unit mass in cgs, \emph{Bottom}: number rate per baryon)} obtained using a
leakage scheme with the diffusion time prescription from \citet{ILEAS2019} and
light bulb absorption (\emph{Left}: \texttt{FLASH} New), \emph{Center}: SedonuGR
Monte-Carlo transport, and the leakage scheme plus light bulb absorption
previously implemented in \citet{MF14} (\emph{Right}: \texttt{FLASH} Old).  The
background fluid quantities correspond to a snapshot of the FLASH New simulation
at $0.6\,\mathrm{ms}$ The new leakage scheme no longer artificially suppresses
the source terms in the midplane of the torus, coming closer to the results of
SedonuGR. Contours of $10^5$g\,cm$^{-3}$ in density are shown as {thick}
dashed lines, corresponding to the density cutoff at which SedonuGR no longer
performs neutrino calculations.}
\label{fig:Comparison06ms}
\end{figure*} 

\begin{figure}
\includegraphics[width=\columnwidth]{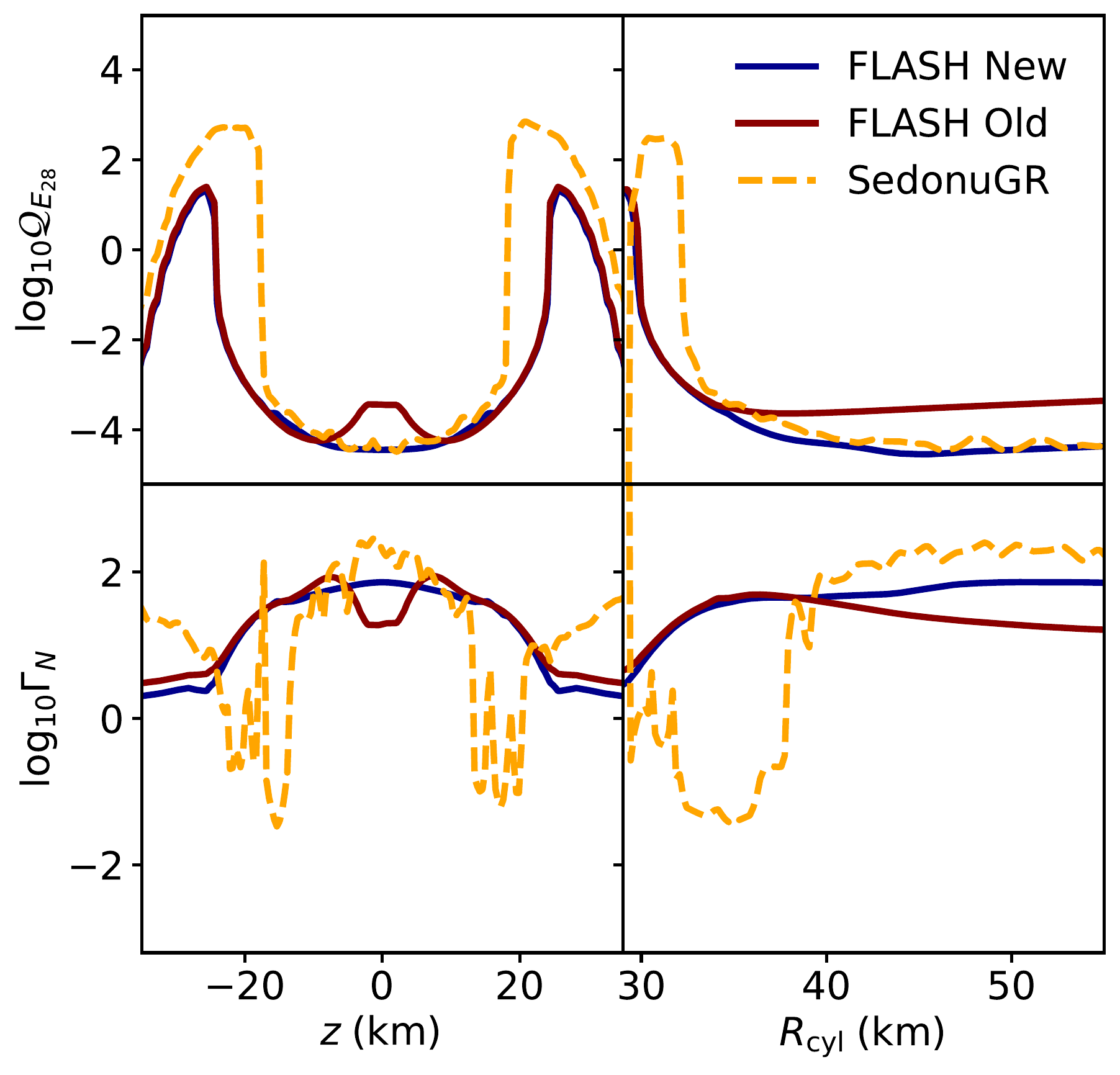}
\caption{Comparison of neutrino source terms in the updated leakage scheme
(\texttt{FLASH} New), SedonuGR, and the leakage scheme previously implemented
(\texttt{FLASH} Old). Like in Figure~\ref{fig:Comparison06ms}, the top rows show
rate of change of internal energy, but this time taking a slice through the
density maximum in the $z$-direction (left), and a slice through the equatorial
plane (right) {of the $0.03\msun$ torus at $0.6\,\mathrm{ms}$}. The
bottom panels are identical slices but for the rate of change of lepton number.
The leakage schemes follow approximately the correct curve, as previously shown
\citep{Richers2015}, but once again the new scheme is closer to the MC method of
SedonuGR, particularly when it comes to the source terms in the torus midplane.}
\label{fig:ComparisonSlices06ms}
\end{figure} 

\begin{figure*}
\includegraphics[width=\textwidth]{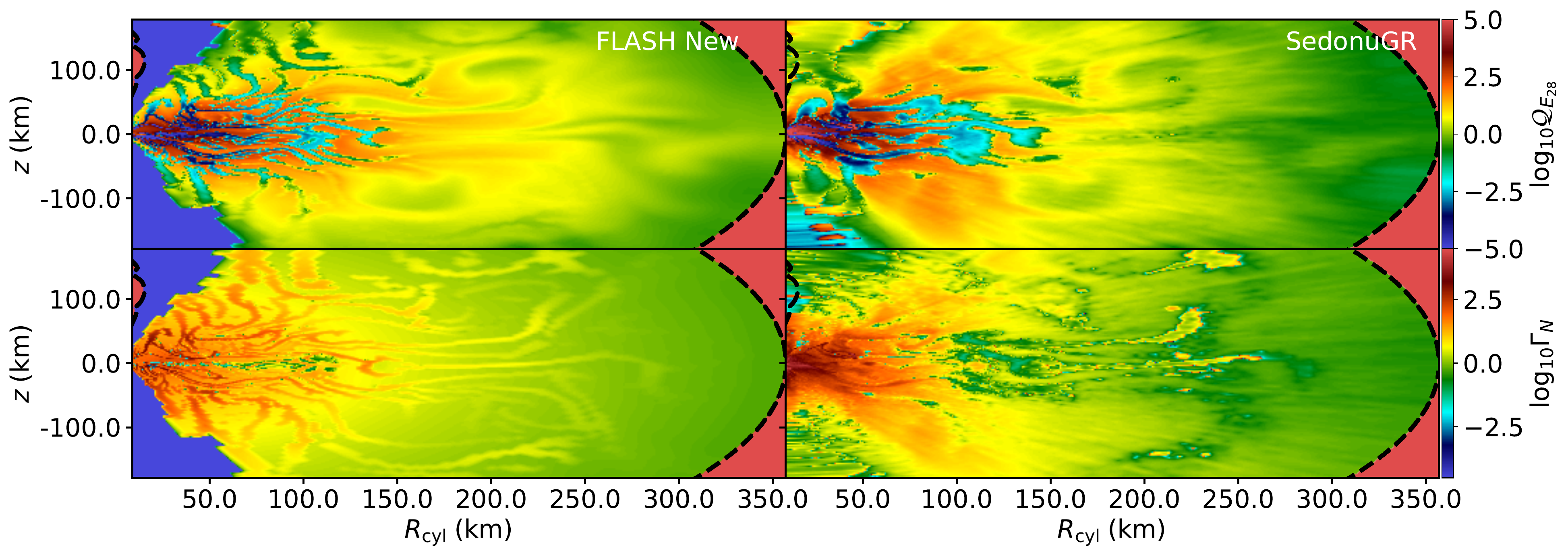}
\caption{Same as Figure~\ref{fig:Comparison06ms}, but for a $0.1\msun$ torus
evolved with 2D-MHD at $\sim 30\,\mathrm{ms}$ {using the
updated leakage scheme and comparing with SedonuGR}.
There are no previous MHD results to compare with. Overall,
SedonuGR and the new leakage scheme show similar trends in both heating/cooling
and change in lepton number. SedonuGR does not discriminate between ambient and
torus material, while our scheme does not perform neutrino calculations on
ambient matter. This appears as purple areas in the poloidal regions of the
\texttt{FLASH} simulations.} 
\label{fig:Comparison30msMt0pt1}
\end{figure*} 

\begin{figure}
\includegraphics[width=0.5\textwidth]{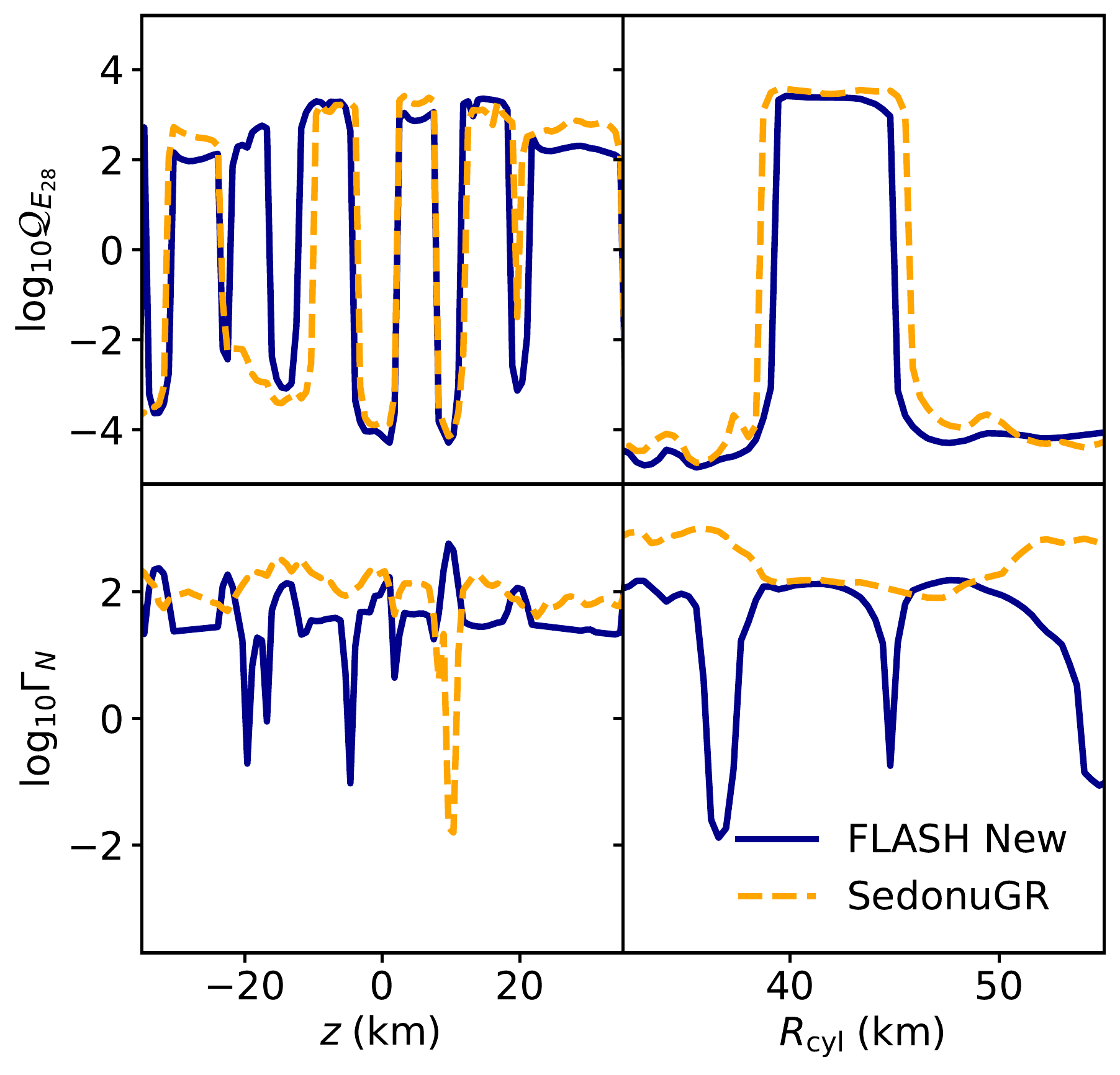}
\caption{Same as Figure~\ref{fig:ComparisonSlices06ms}, but for a $0.1\msun$
torus evolved with 2D-MHD {and the updated leakage scheme} at
$\sim 30$\,ms. There are no previous MHD results to compare with, but the overall
effects of neutrinos are captured well, with significant deviations only in
small localized regions.}  
\label{fig:ComparisonSlices30msMt0pt1}
\end{figure} 

\subsection{Implementation in \tt{FLASH4.5}} 
We extend the leakage and light bulb absorption scheme of \citet{FM13} and \citet{MF14} by
computing the diffusion time with equation~(\ref{eq:tdiffEILEAS}) and implement
it in {\tt FLASH4.5}.  Neutrino energy and number gradients are obtained using
second order finite differences in each spatial direction,
The flux limiters are calculated for each species as in (\ref{eq:FLDLimE}). The
flux limiters are then combined with the gradients, energy/number densities, and
opacities to form the fluxes in \ref{eq:FicksE}. The fluxes are then linearly
interpolated to the cell faces, such that we can take a numerical divergence
using Gauss's theorem (\ref{eq:GaussLaw}) and the face areas of a given cell,
analogous to the unsplit update in \S\ref{sec:MHDOverview}. 

\subsubsection{Comparison with 2D hydrodynamic simulations} 
We test the effect of the new diffusion time on the long-term disk evolution by
running a 2D axisymmetric hydrodynamic simulation of an accretion torus using
the previous leakage scheme and the new one. A full description of the
methodology is available in \citet{FM13,MF14}, here we briefly describe the
initial setup.  The compact object is a BH with mass $3 \msun$, and the orbiting
equilibrium torus is chosen to be optically thin with a mass of $0.03 \msun$.
The entropy, electron fraction, and torus distortion parameter are chosen to be
$8 k_B/\mathrm{baryon}$, $0.1$, and $1.911$, respectively. This yields a
well-studied initial condition compatible with dynamical merger simulations
\citep{FM13,MF14,Richers2015,Lippuner2017}.  Angular momentum transport is
handled with the viscous stress parameterization of \citet{Shakura1973}, with
$\alpha$ set to $0.05$. The tori are evolved for 1 second.

The average mass-flux-weighted electron fraction of the outflow increases by
$5\%$ when using the new diffusion time formulation, as shown in the histogram
of torus outflows in Figure~\ref{fig:FLASH3TorusComparison}. This
{result corresponds} to a decrease in
diffusion time by a factor of $\sim10$ in the high-density regions of the torus.
Neutrinos are still preferentially trapped near the midplane, but the new
methodology no longer overestimates this effect for the optically thin torus.
The new method yields about 50\% less mass outflows, predominately due to the
same effect: the energy deposited into early time outflows is cooled more
efficiently, unbinding less material. The major difference occurs in the
neutrino-driven outflow ejected from the back of the torus, with electron
fraction $Y_e\sim0.25$. We caution that this difference will likely not extend
to the late-time, radiatively inefficient evolution, as the energy and density
of late time outflows appear identical, and affects mainly the initial winds. 

\subsubsection{Comparison with SedonuGR in Hydro} 
We compare the emission and absorption of neutrinos to the Monte-Carlo neutrino
transport code SedonuGR \citep{Richers2015,Richers2017}. We are interested in
the rate of change of lepton number, which governs the change in electron
fraction, and the rate of change of internal energy, which governs changes in
internal energy of the ejecta due to neutrinos. Following \citet{Richers2015},
we load a fluid snapshot from the \texttt{FLASH4.5} simulation using the new
method in SedonuGR. We use the Helmholtz EOS in SedonuGR for
consistency.

At late times, the neutrino scheme for the optically thin torus shows minimal
differences with the full Monte-Carlo results \citep{Richers2015}. The changes
to the neutrino scheme are most evident at very early times, so we choose a time
of $\sim$0.6\,ms for the comparison, the result of which is shown in
Figure~\ref{fig:Comparison06ms}. Most of the differences in source terms occur
in the regions where absorption becomes important, which in our approach is
handled by an approximate light bulb implementation \citep{FM13,MF14}, which we
did not modify here. Importantly, the new neutrino leakage scheme no longer
suppresses neutrino emission in the midplane of the torus, where the pressure
scale height is comparatively large. This is shown quantitatively with slices in
the equatorial plane and along the $z$-direction at the torus density maximum in
Figure~\ref{fig:ComparisonSlices06ms}. 

\subsubsection{MHD Comparison with SedonuGR}
Since our base torus is more massive than those used in the previous
hydrodynamic test runs, and MHD evolution differs in comparison to viscous
hydrodynamics, we also show a comparison of our leakage scheme in
2D-MHD with results from SedonuGR (Figure~\ref{fig:Comparison30msMt0pt1}). Equatorial and vertical slices
through the density maximum of a torus identical to our base model are shown in
Figure~\ref{fig:ComparisonSlices30msMt0pt1} at 30\,ms, when the neutrinos are
important for setting the electron fraction of the outflows. Importantly, the
source term modifying the electron fraction remains {of} the
same order of magnitude across most of the torus.

\bsp
\label{lastpage}
\end{document}